\documentclass[preprint]{aastex63}

\newcommand\HHO{H$_2$O }
\newcommand\KGMS{kg m$^{-2}$ s$^{-1}$}
\newcommand\KMS{km s$^{-1}$}

\usepackage{enumitem,amssymb}
\usepackage{amsmath}
\usepackage{pifont}
\usepackage{multirow}

\graphicspath{ {JPGs/} }
\received{December 11, 2020}
\revised{March 15, 2021}
\accepted{March 16, 2021}

\acceptjournal{PSJ}

\shorttitle{Meteoroids at Ceres}
\shortauthors{Pokorn\'{y} et al.}

\begin{document}

\title{Erosion of volatiles by micro-meteoroid bombardment on Ceres, and comparison to the Moon and Mercury}

\correspondingauthor{Petr Pokorn\'{y}}
\email{petr.pokorny@nasa.gov, pokorny@cua.edu}

\author[0000-0002-5667-9337]{Petr Pokorn\'{y}}
\affiliation{The Catholic University of America, 620 Michigan Ave, NE Washington, DC 20064, USA}
\affiliation{Goddard Space Flight Center, 8800 Greenbelt Rd., Greenbelt, MD, 20771, USA}

\author[0000-0003-3456-427X]{Erwan Mazarico}
\affiliation{Goddard Space Flight Center, 8800 Greenbelt Rd., Greenbelt, MD, 20771, USA}

\author[0000-0002-5821-4066]{Norbert Schorghofer}
\affiliation{Planetary Science Institute, Tucson, AZ 85719, USA}





\begin{abstract}
(1) Ceres, the largest reservoir of water in the main-belt, was recently visited by the Dawn spacecraft that revealed several areas bearing H$_2$O-ice features. Independent telescopic observations showed a water exosphere of currently unknown origin. We explore the effects of meteoroid impacts on Ceres considering the topography obtained from the Dawn mission using a \replaced{full-fledged}{widely-used} micro-meteoroid model and ray-tracing techniques. Meteoroid populations with 0.01-2~mm diameters are considered. We analyze the short-term effects Ceres experiences during its current orbit as well as long-term effects over the entire precession cycle. We find the entire surface is subject to meteoroid bombardment leaving no areas in permanent shadow with respect to meteoroid influx. The equatorial parts of Ceres produce 80\% more ejecta than the polar regions due to the large impact velocity of long-period comets. Mass flux, energy flux, and ejecta production vary seasonally by a factor of \replaced{a few}{3--7} due to the inclined eccentric orbit. Compared to Mercury and the Moon, Ceres experiences significantly smaller effects of micro-meteoroid bombardment, with a total mass flux of $4.5\pm1.2\times10^{-17}$~\KGMS. On average Mercury is subjected to a $50\times$ larger mass flux and generates $700\times$ more ejecta than Ceres, while the lunar mass flux is $10\times$ larger, and the ejecta generation is $30\times$ larger than on Ceres. For these reasons, we find that meteoroid impacts are an unlikely candidate for the production of a water exosphere or significant excavation of surface features. The surface turnover rate from the micro-meteoroid populations considered is estimated to be 1.25~Myr on Ceres.
\end{abstract}


\section{Introduction} \label{SEC:INTRO}
Ceres is the only dwarf planet with a global high-resolution shape model thanks to the Dawn mission and the Dawn spacecraft's Framing Camera\footnote{\url{https://sbnarchive.psi.edu/pds3/dawn/fc/DWNCSPC_4_01/}}. The dwarf planet itself is surrounded by the zodiacal cloud, \added{a cloud of dust and meteoroids enveloping the entire solar system}, that is mainly sourced from main-belt asteroids and short/long-period comets \citep{Nesvorny_etal_2010}. \added{Since the zodiacal cloud is increasingly denser with decreasing heliocentric distance \citep{Leinert_etal_1981}, Ceres is expected to experience a smaller meteoroid flux and impact velocities than those seen by the Moon and Mercury.}

Two sets of telescopic observations suggest that Ceres has at least temporarily a water exosphere \citep{Ahearn_Feldman_1992, Kuppers_etal_2014}. Despite negative results from follow-up observations with different observational facilities \citep{rousselot2011,roth2016,mckay2017,Roth_2018,rousselot2019}, the production of the water exosphere on Ceres was the subject of many works \citep{tu2014,Schorghofer_etal_2016,formisano16,Landis_etal_2017,Villarreal_etal_2017, Schorghofer_etal_2017}.  
\cite{Landis_etal_2019} has analyzed the surface evolution based on the flux of \replaced{larger impactors}{impactors larger than 100m in diameter}, however the effects of smaller meteoroids are yet to be assessed. 
Therefore, it is natural to investigate this gap and address how different meteoroid populations imprint their activity on the surface of Ceres, how they are affected by the complex topography, and how they influence the regions with exposed water-ice.

Until recently, the availability of precise shape models and detailed dynamical meteoroid models was scarce; to say nothing of the existence of models to study the effects of meteoroid bombardment using ray-tracing techniques.
\citet{Pokorny_etal_2020} analyzed the effects of meteoroid bombardment on the lunar surface, specifically on both lunar poles using detailed topography derived from Lunar Orbiter Laser Altimeter (LOLA) observations \citep{Smith_etal_2010}. 
Advanced solar illumination models \citep{Mazarico_etal_2018} show that the lunar poles harbor many cold traps and permanently shadowed regions \added{(PSRs)}, consistent with spacecraft data \citep[e.g., Diviner instrument][]{Paige_etal_2010}. On the other hand, \citet{Pokorny_etal_2020} showed that meteoroids, due to their broad range of impact directions, reach even the deepest craters on both lunar poles, including those that are permanently shadowed from the Sun.
\citet{Pokorny_etal_2020} also showed that the crater walls, that \replaced{are}{were} more inclined toward the high energetic meteoroid flux concentrated close to the ecliptic, \replaced{experience}{experienced} a higher rate of impact gardening compared to the crater floors. The floors \replaced{are}{were} partially shielded from the meteoroid flux and \replaced{experience}{experienced} the high energetic impacts at grazing angles.

Compared to the Moon, Ceres is on an inclined, eccentric orbit, embedded inside the source region of a significant portion of the zodiacal dust cloud \citep[main-belt asteroids; e.g.][]{Nesvorny_etal_2006}. 
The obliquity of Mercury and the Moon is stable in the long term \citep{goldreich66}, which allows the existence of long-lived permanently shadowed environments, whereas the obliquity of Ceres oscillates between $\epsilon=2^\circ$ and $\epsilon=20^\circ$ with a period of 24.5 kyr and complicates the existence of permanently shadowed regions \deleted{(PSRs)} \citep{Ermakov_etal_2017}. 
However, as shown in \citet{Ermakov_etal_2017}, PSRs indeed exist on Ceres and can potentially retain water-ice. In addition to the ice deposits in the PSRs, \citet{Combe_etal_2019} showed nine areas with \HHO absorption features detected in Dawn VIR spectra ($2.0 \mu$m line). These areas are not currently accessible by direct solar radiation, however, as shown in \citet{Pokorny_etal_2020} they might be accessible to meteoroid impacts. \added{These meteoroid impacts bring exogenous energy to PSRs that can remove volatiles from PSRs. It has been suggested that destabilized volatiles might produce a tenuous exosphere at airless bodies such as Mercury and the Moon \citep{Cintala_1992}}, which motivates us to quantify the effect of meteoroids on these interesting regions at Ceres.

\citet{Landis_etal_2019} discussed various explanations for the temporary existence of the water exosphere on Ceres: (a) sublimation from sub-surface water-ice tables \citep{fanale1989,Prettyman_etal_2017}; (b) sublimation from transient surface exposures of water-ice \citep{Landis_etal_2017}; (c) sputtering by solar energetic particle events from surface ice \citep{Villarreal_etal_2017}; (d) seasonal, optically thin water ice polar deposit \citep{Schorghofer_etal_2017}. The telescopic evidence from \citet{Ahearn_Feldman_1992, Kuppers_etal_2014} suggests 3 -- 6 kg s$^{-1}$ of water vapor produced to sustain the observed exosphere. None of these mechanisms or their combination provides such a high rate \citep{Landis_etal_2019}. The residence time of the water exosphere was shown to be around 7~hr \citep{Schorghofer_etal_2016}, thus a continuous active source is needed to sustain the longer lasting exosphere observed by \citet{Kuppers_etal_2014}. Since meteoroids provide a quasi-continuous source of energy to surfaces of airless bodies, we aim to estimate the meteoroid impact driven production rates and provide a missing piece to this puzzle. Moreover, meteoroids and dust may also erode water ice deposits. 

\section{Methods}

\subsection{Ceres topography model}
For the representation of Ceres, we use an object (OBJ) file containing 1,579,014 vectors and 3,145,728 faces (triangles) derived from the digital terrain model (DTM) constructed from Framing Camera 2 (FC2) images taken during Dawn High Altitude Mapping Orbit (HAMO). HAMO DTM covers approximately 98\% of the cererean surface with a lateral spacing of $\approx136.7$ m/pixel\footnote{The mission data product can be found at \url{https://sbnarchive.psi.edu/pds3/dawn/fc/DWNCSPC_4_01/DATA/ICQ/CERES_SPC181019_0512.ICQ}}. We converted the original ICQ file to OBJ file using an example code described here \url{https://sbnarchive.psi.edu/pds3/dawn/fc/DWNCSPC_4_01/DOCUMENT/ICQMODEL.ASC}. We also provide the working version of this code at the project's GitHub page (see Software section). 

\subsection{Ray-tracing code}
\label{SEC:RAY-TRACING}
We improved our own procedure from \citet{Pokorny_etal_2020} \added{in terms of computational speed} and combined it with two external libraries \added{written in C++}: (1) \texttt{tinyobjloader} for loading up to 10 million polygon models and (2) \texttt{fastbvh} - a Bounding Volume Hierarchy algorithm that uses axis-aligned bounding box (AABB) trees to efficiently calculate line-triangle intersections. Our program loads the OBJ file using \texttt{tinyobjloader} into memory and then constructs using \texttt{fastbvh}, an AABB tree that allows us to quickly evaluate ray-face intersections for the entire triangular mesh. For each combination of longitude, latitude, and velocity in the meteoroid model our code determines whether each surface triangle is reachable or obstructed/shadowed by any other triangle.
We also calculate the incidence angle $\varphi$ for each surface triangle and meteoroid direction/velocity combination independently. Each surface triangle represents a plane for which we calculate a normal vector pointing outside of the object, $\overrightarrow{n_\mathrm{sur}}$. Then  $\cos{\varphi}=-\overrightarrow{n_\mathrm{sur}} \cdot \overrightarrow{e_\mathrm{imp}}$, where $\overrightarrow{e_\mathrm{imp}}$ is the velocity vector of the impacting meteoroid normalized to unity. For impacts perpendicular to the surface $\cos{\varphi}=1$, whereas for meteoroids at grazing angles $\cos{\varphi} \rightarrow 0$, since $\varphi \rightarrow 90^\circ$.

\subsection{Model for meteoroid environment at Ceres and its variations over current Ceres' orbit}\label{SEC:Meteoroid_Model}
\begin{figure}
\epsscale{1.1}
\plotone{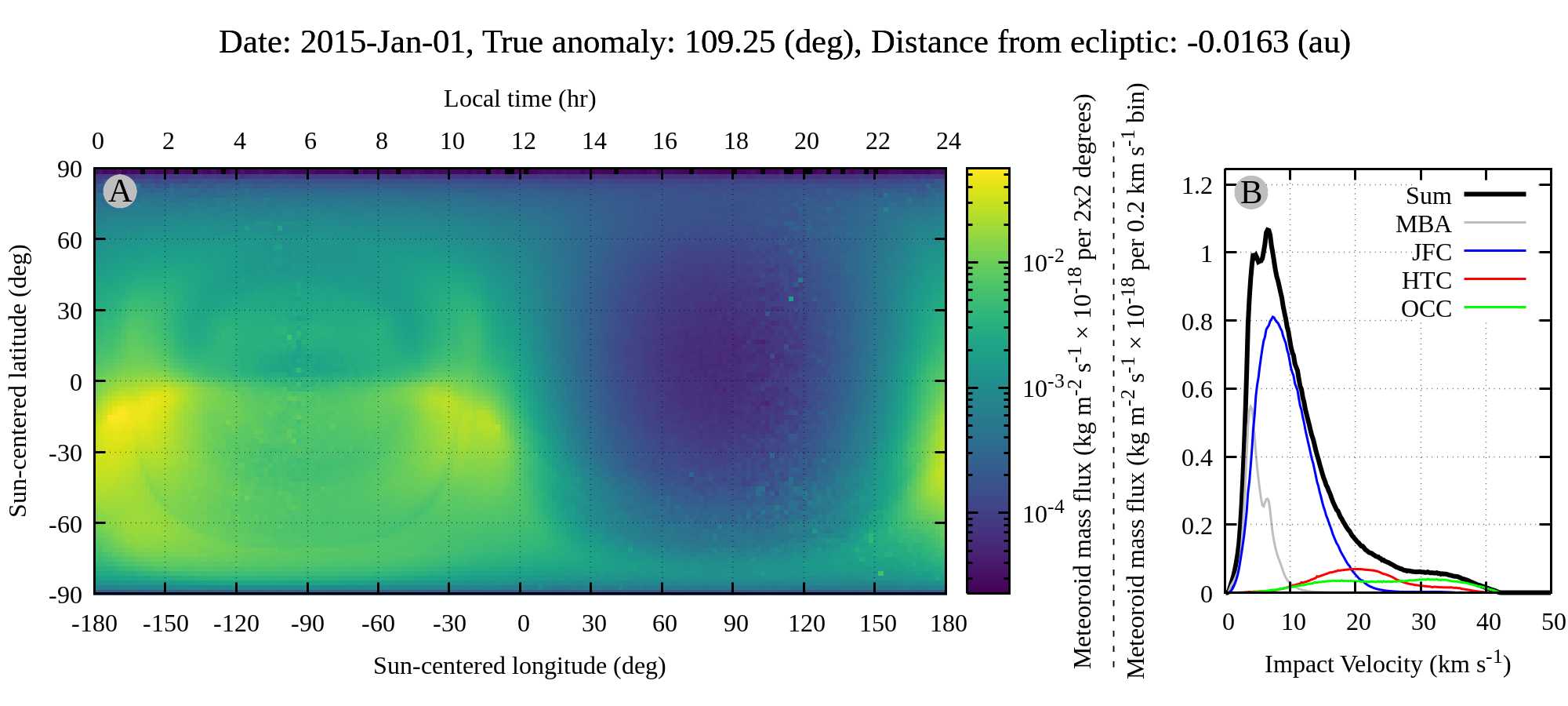}
\caption{
\label{FIG:RADIANT+VEL}
Panel (A): Distribution of the meteoroid mass flux on the celestial sphere as seen from Ceres on January 1st, 2015. The $x$-axis shows the sun-centered longitude, where the origin ($0^\circ,0^\circ$) is the direction of the Sun, while the y-axis is calculated from Ceres' orbital plane ($\beta = 0^\circ$), where $\beta=\pm90^\circ$ are impacts with relative velocity vectors perpendicular to the orbital plane. The color bar is a logarithmic scale in grams per second per 2 by 2 degrees in longitude and latitude. Meteoroids impacting from the ram direction ($-180^\circ<\lambda-\lambda_\odot<0^\circ$) dominate the mass flux throughout the entire orbit. Panel (B): Histogram of the meteoroid mass flux with respect to the impact velocity (0.2 km s$^{-1}$ bins). Different color lines show the contribution of the four meteoroid populations: MBAs, JFCs, HTCs, and OCCs, while the black line represents their sum. Movie showing one full orbit is available in the supplementary materials.
}
\end{figure}

The meteoroid model in this work uses the same constraints and configuration as the model used in \citet{Pokorny_etal_2018,Pokorny_etal_2019,Pokorny_etal_2020} for studies of Mercury's and the Moon's meteoroid environments. \added{This means we do not change the configuration of the zodiacal cloud/meteoroid model with respect to those previous studies, ensuring they are comparable with each other}. Our model combines contributions of four meteoroid populations originating from main-belt asteroids (MBAs), Jupiter-Family comets (JFCs), Halley-type comets (HTCs), and Oort Cloud comets (OCCs). These four populations dominate the meteoroid mass and number density flux in the inner solar system \citep{Nesvorny_etal_2010,Nesvorny_etal_2011JFC}, while the outer solar system is mostly dominated by Edgeworth-Kuiper Belt (EKB) meteoroids \citep{Poppe_etal_2019}. The details of the model used here are summarized in Table \ref{TABLE:Model_Description} and references therein. We \replaced{calculated}{calculate} the distribution of directions and velocities of impacting meteoroids for one orbit from January 1$^{st}$, 2015 to August 18$^{th}$, 2019 in 10-day intervals resulting in 169 individual snapshots of the meteoroid environment at Ceres. Each of these snapshots provides the meteoroid mass flux distributed in sun-centered longitudes $\lambda-\lambda_\odot$ and latitudes $\beta$ with 2 degree resolution (i.e., directions) and impact velocities with 2 km s$^{-1}$ resolution, thus providing the full three-dimensional map of velocity vectors for meteoroids in each meteoroid population separately. An example of the meteoroid environment map and the impact velocity distribution is shown in Figure \ref{FIG:RADIANT+VEL}. The directionality of meteoroids on Ceres is similar to that seen on other airless bodies like Mercury \citep{Pokorny_etal_2018} or the Moon \citep{Pokorny_etal_2019}. The mass flux is dominated by meteoroids impacting Ceres from directions close to the orbital plane $(\beta \approx 0^\circ$) and within $90^\circ$ of the ram/apex direction $(-180^\circ < \lambda - \lambda_\odot < 0^\circ)$,
where JFC meteoroids dominate the total influx ($-180^\circ<\lambda-\lambda_\odot<-140^\circ, -30^\circ<\beta<30^\circ $ and $-40^\circ<\lambda-\lambda_\odot<-0^\circ, -30^\circ<\beta<30^\circ$). HTC and OCC meteoroids preferentially originate from the apex direction (a circle around [$\lambda-\lambda_\odot,\beta]=[-90^\circ,0^\circ]$ with $45^\circ$ radius). MBA meteoroids have the smallest relative impact velocities $V_\mathrm{imp}<10$ km s$^{-1}$ and impact Ceres preferentially from higher Sun-centered latitudes $|\beta|>60^\circ$ and/or from the sun/anti-sun directions ($\lambda-\lambda_\odot \approx -180^\circ$ and $\lambda-\lambda_\odot \approx 0^\circ$). The impact velocity distribution (Figure \ref{FIG:RADIANT+VEL}B) shows that MBA meteoroids have a median $V_\mathrm{imp_\mathrm{50\%}} = 5.5$ km s$^{-1}$, JFC meteoroids are slightly faster with $V_\mathrm{imp_\mathrm{50\%}}= 9.2$  km s$^{-1}$, whereas the long-period comet sources provide much more energetic impactors with $V_\mathrm{imp_\mathrm{50\%}}=20.5$  km s$^{-1}$ for HTC meteoroids and $V_\mathrm{imp_\mathrm{50\%}}= 25.3$ km s$^{-1}$ for OCCs meteoroids.

\begin{figure}
\epsscale{1.1}
\plotone{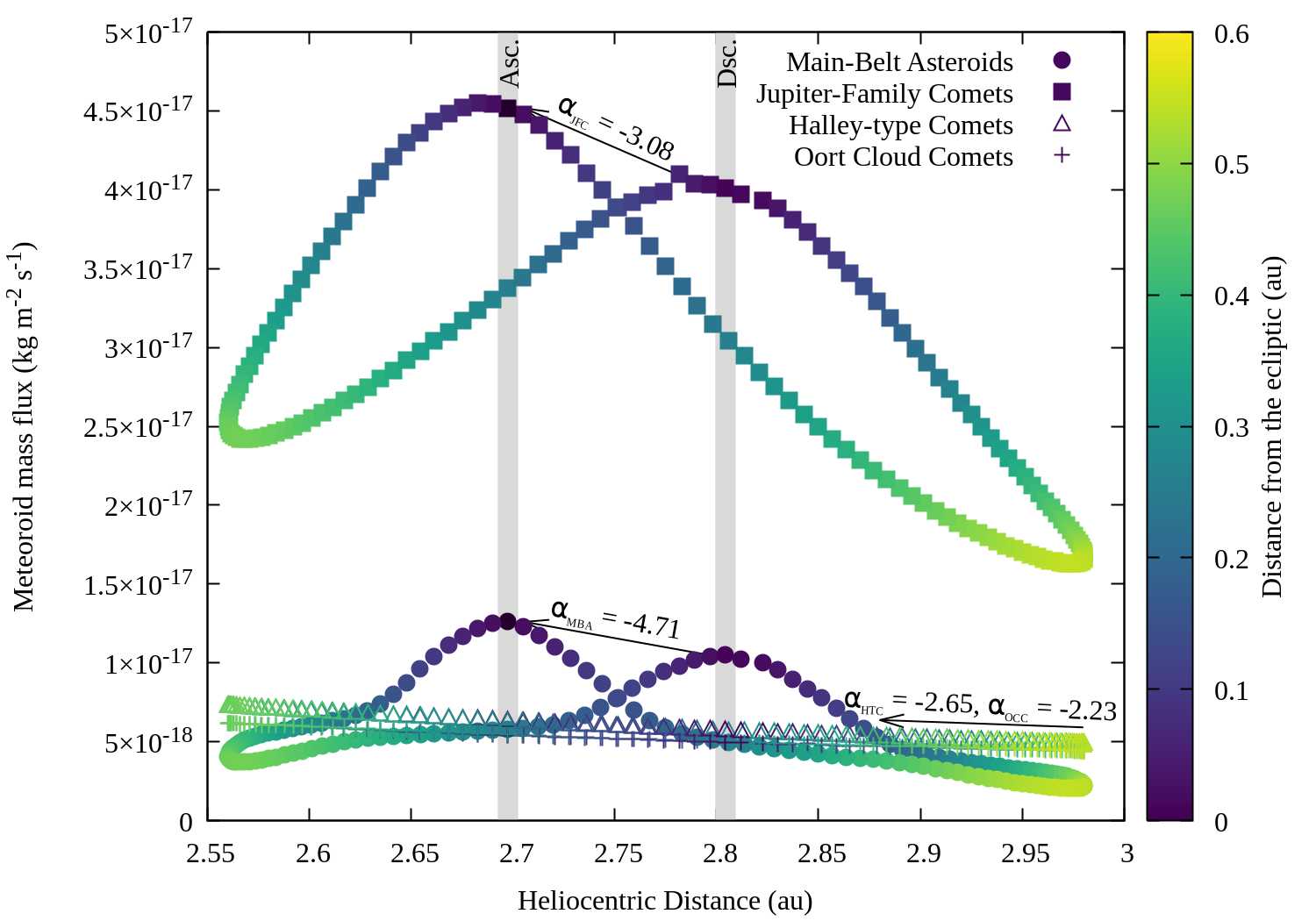}
\caption{
\label{FIG:FLUX_vs_Orbit}
Variations of the meteoroid mass flux at Ceres with the heliocentric distance for four meteoroid populations analyzed in this paper. Each of the populations is color coded by Ceres' distance from the ecliptic. Meteoroids originating in MBA and JFC populations show a strong correlation of the meteoroid mass flux and the distance from the ecliptic. HTC and OCC meteoroids are unaffected by the distance from the ecliptic. Arrows and their corresponding labels show the scaling proportionality of the meteoroid mass flux $\mathcal{M}$ with the heliocentric distance $r$, $\mathcal{M}\propto r^{\alpha}$. Grey areas show the moments of Ceres' ecliptic crossings that correspond to two spikes in $\mathcal{M}$ for MBA and JFC meteoroids. 
}
\end{figure}

Due to the non-zero eccentricity and inclination of Ceres, the meteoroid environment undergoes significant changes during one orbit (Fig. \ref{FIG:FLUX_vs_Orbit}). During one of its orbital cycles, Ceres crosses the ecliptic twice (gray regions in Fig. \ref{FIG:FLUX_vs_Orbit}); this is where we record the global and local maxima of the meteoroid mass flux $\mathcal{M}$ for MBA and JFC meteoroids. This is not unexpected because MBA and JFC meteoroid models show that they are concentrated close to the ecliptic \citep{Nesvorny_etal_2010, Nesvorny_etal_2011JFC}. On the other hand, HTC and OCC meteoroids are unaffected by Ceres' departure from the ecliptic plane due to the broad range of inclinations of their parent bodies \citep{Nesvorny_etal_2011OCC, Pokorny_etal_2014}. Overall, JFC meteoroid mass flux dominates the total mass influx at Ceres throughout the entire orbit due to the dominance of JFC meteoroid in the Zodiacal cloud.  This stems from the population mixing ratios obtained from \citet{CarrilloSanchez_etal_2016} and \citet{Pokorny_etal_2019}, where JFCs dominate the terrestrial flux by a factor of 10 compared to other meteoroid populations. The departure from the ecliptic is not the only factor shaping the meteoroid mass flux at Ceres. From spacecraft observations and modelin, we know that the Zodiacal Cloud density inside 1 au increases with heliocentric distance, $r$ \citep[e.g, ][]{Leinert_etal_1981,Pokorny_etal_2019}, while the outer portions of the Zodical Cloud show constant density \citep{Poppe_etal_2019}. With Ceres being inside the main-belt, we can expect significant differences from what \replaced{was}{is} observed inside 1 au for MBA and JFC meteoroids. For the outer solar system sources (HTC and OCC), the meteoroid mass flux scaling should be similar to that observed/modeled inside 1 au. We \replaced{calculated}{calculate} the proportional changes of $\mathcal{M}$ for all four sources assuming the single power-law scaling $\mathcal{M} \propto r^\alpha$ and positions close to the ecliptic $|z|<0.01$ au (labels above arrows in Fig. \ref{FIG:FLUX_vs_Orbit}). While not directly comparable to the Zodiacal Cloud density, due to the velocity changes that Ceres experiences during its orbit, the proportionality of HTC ($\propto r^{-2.65}$) and OCC ($\propto r^{-2.23}$) meteoroids is quite similar to those modeled in \citet{Pokorny_etal_2019}. On the other hand, JFC ($\propto r^{-3.08}$) and most significantly MBA ($\propto r^{-4.71}$) meteoroids show much steeper scaling with heliocentric distance than what was observed inside 1 au. \added{The two main factors that drive the proportionality of different populations are: 1) the location of the source region with respect to the target (Ceres) where MBA and JFC meteoroids are ejected on semimajor axes close to that of Ceres while long-period comet meteoroids can be considered distance source regions. This causes larger variations for the close source regions due to the fact that some meteoroids are released at semimajor axes smaller than that of Ceres; and 2) the impact velocity of different populations at Ceres which is modulated by the populations' semimajor axis/eccentricity/inclination distributions. MBA and JFC meteoroids can acquire very small impact velocities at Ceres due to their orbital similarity, while for the long-period meteoroids this is insignificant}.

We \replaced{calculated}{calculate} that, averaged over one of its current orbits around the Sun, the mean meteoroid flux at Ceres is $\overline{\mathcal{M}}=4.49\pm1.18\times
10^{-17}$ \KGMS, which is approximately 9.5 times smaller than that at the Moon  $\overline{\mathcal{M}_\mathrm{Moon}}=42.17\times 10^{-17}$ \KGMS\ and 22 times smaller than the terrestrial meteoroid mass flux $\overline{\mathcal{M}_\mathrm{Earth}}=98.53\times 10^{-17}$ \KGMS. The mean contributions of the four meteoroid populations are the following: $\overline{\mathcal{M}_\mathrm{MBA}}=0.52\pm0.26\times 10^{-17}$ \KGMS, $\overline{\mathcal{M}_\mathrm{JFC}}=2.85\pm0.86\times 10^{-17}$ \KGMS, $\overline{\mathcal{M}_\mathrm{HTC}}=0.60\pm0.08\times 10^{-17}$ \KGMS, $\overline{\mathcal{M}_\mathrm{OCC}}=0.52\pm0.06\times 10^{-17}$ \KGMS. The median impact velocities averaged throughout the entire orbit are the  following: $\overline{V_\mathrm{50\%}}\mathrm{(MBA)}=4.73\pm0.69$ \KMS, $\overline{V_\mathrm{50\%}}\mathrm{(JFC)}=9.32\pm0.46$ \KMS, $\overline{V_\mathrm{50\%}}\mathrm{(HTC)}=20.47\pm0.63$ \KMS, $\overline{V_\mathrm{50\%}}\mathrm{(OCC)}=25.31\pm1.05$ \KMS. The overall shape of population velocity distributions shown in Fig. \ref{FIG:RADIANT+VEL} does not significantly change throughout the orbit, only their relative contributions that consequently change the overall velocity distribution at Ceres.
%


\begin{deluxetable}{llcl|l}

\tablecaption{\label{TABLE:Model_Description}Description of meteoroid dynamical models used in this work. The meteoroid model used here has six free parameters: the collisional lifetime multiplier $F_\mathrm{coll}$ \citep{Pokorny_etal_2014}, differential size-frequency index $\alpha$, and the average daily mass influx at Earth $M$ for each of the four populations in metric tons per day (1000 kg per day or 11.57 g s$^{-1}$) \citep{CarrilloSanchez_etal_2016,Pokorny_etal_2019}. For more detailed information refer to references in the table or \citet{Pokorny_etal_2018}.}

\tablenum{1}

\tablehead{\colhead{Source population} & \colhead{Acronym} & \colhead{Diameter {($\mu$m)}} & \colhead{Reference} & \colhead{Parameter Settings} } 


\startdata
Main-belt asteroids & MBA & 10 -- 2000 & \citet{Nesvorny_etal_2010} & $F_\mathrm{coll}=20$, $\alpha=-4.0$  \\ \cline{5-5}
Jupiter-family comets & JFC & 10 -- 2000 & \citet{Nesvorny_etal_2011JFC} & Mass influx at Earth (tons/day) \\
Halley-type comets & HTC & 10 -- 2000 & \citet{Pokorny_etal_2014}& $M_\mathrm{MBA}=3.7$, $M_\mathrm{HTC}=2.82$  \\
Oort Cloud comets & OCC & 10 -- 2000 & \citet{Nesvorny_etal_2011OCC} & $M_\mathrm{JFC}=34.6$, $M_\mathrm{OCC}=2.12$ \\ 
\enddata




\end{deluxetable}

\subsection{\label{SEC:Definitions}Quantities induced by the meteoroid bombardment}

In this paper we discuss four different quantities that result from the bombardment of the Cererean surface by interplanetary meteoroids: (1) the meteoroid mass flux $\mathcal{M}$, (2) the meteoroid energy flux $\mathcal{E}$, (3) the ejecta mass produced by impacting meteoroids $\mathcal{P}^+$, and (4) the area of craters produced by meteoroid bombardment $\mathcal{A}$.

In Section \ref{SEC:RAY-TRACING} we \replaced{described}{describe} how we determine whether each surface element is reachable by the flux of meteoroids incoming for a selected direction (i.e., it is not shadowed by any feature on the surface of the dwarf planet). This analysis results in a list of $\cos{\varphi}$ values \added{(cosines of incident angles)} for each triangular element on Ceres, and the value $S(\lambda-\lambda_\odot,\beta)$ which represents the percentage of shadowing of a particular element ($S=0$ is a completely shadowed element, $S=1$ is unobstructed). Here, $\lambda-\lambda_\odot$ and $\beta$ are the sun-centered longitude and latitude, i.e., the angles representing the meteoroid directionality. \added{The meteoroid model we use here gives us the five-dimensional information for each time snapshot we analyze in this manuscript: for each combination of directions $\lambda-\lambda_\odot$, impact velocity $v_\mathrm{imp}$, and meteoroid diameter $D$ we obtain the meteoroid number flux $N_\mathrm{met}$. Assuming that all meteoroids are spheres, we get the meteoroid mass flux $M_\mathrm{met}$ and cross-sectional area of impacting meteoroids per unit time $A_\mathrm{met}$ as:
\begin{eqnarray}
M_\mathrm{met}(\lambda-\lambda_\odot,\beta, v_\mathrm{imp}) &=& \sum_{D}  \frac{\pi}{6}D^3\rho_\mathrm{met} N_\mathrm{met}(\lambda-\lambda_\odot,\beta, v_\mathrm{imp},D),\\
A_\mathrm{met}(\lambda-\lambda_\odot,\beta, v_\mathrm{imp}) &=& \sum_{D}  \frac{\pi}{4}D^2 N_\mathrm{met}(\lambda-\lambda_\odot,\beta, v_\mathrm{imp},D),
\end{eqnarray}
where we adopt the meteoroid bulk density $\rho_\mathrm{met} = 2,000$ kg m$^{-3}$ used in meteoroid models we show in Table \ref{TABLE:Model_Description}.}

Equipped with these quantities, we then calculate for each triangular element on Ceres $\mathcal{M}$, $\mathcal{E}$, $\mathcal{P}^+$, and $\mathcal{A}$, defined as follows:
\begin{eqnarray}
\mathcal{M} &=& \sum_{\lambda-\lambda_\odot,\beta, v_\mathrm{imp}} M_\mathrm{met} (\lambda-\lambda_\odot,\beta, v_\mathrm{imp})   S(\lambda-\lambda_\odot,\beta) \cos{\varphi}(\lambda-\lambda_\odot,\beta),\\
    \mathcal{E} &=& \sum_{\lambda-\lambda_\odot,\beta,v_\mathrm{imp}} \frac{1}{2} M_\mathrm{met} (\lambda-\lambda_\odot,\beta,v_\mathrm{imp}) v_\mathrm{imp}^2(\lambda-\lambda_\odot,\beta) S(\lambda-\lambda_\odot,\beta) \cos{\varphi}(\lambda-\lambda_\odot,\beta), \\
    \mathcal{P^+} &=& \mathcal{C}\sum_{\lambda-\lambda_\odot,\beta,v_\mathrm{imp}} M_\mathrm{met} (\lambda-\lambda_\odot,\beta,v_\mathrm{imp}) v_\mathrm{imp}^{2.46}(\lambda-\lambda_\odot,\beta) S(\lambda-\lambda_\odot,\beta) \cos^3{\varphi}(\lambda-\lambda_\odot,\beta), \label{EQ:EJECTA} \\
     \mathcal{A} &=& F_\mathrm{cr} \sum_{\lambda-\lambda_\odot,\beta, v_\mathrm{imp}} A_\mathrm{met} (\lambda-\lambda_\odot,\beta,v_\mathrm{imp})  S(\lambda-\lambda_\odot,\beta) \cos{\varphi}(\lambda-\lambda_\odot,\beta), \label{EQ:Area_Coverage}
\end{eqnarray}
where 
\deleted{$v_\mathrm{imp}$ is the meteoroid impact velocity, and }$\mathcal{C} = 7.358 \mathrm{~km}^{-2} \mathrm{~s}^2$ is a scaling constant determined from the laboratory experiments reported by \citet{Koschny_Grun_2001}.We assume the impactor-to-crater area ratio $F_\mathrm{cr} = 63$, which is based on fitting results from Table 1 in \citet{Koschny_Grun_2001}. {The scaling constant $\mathcal{C}$ describes the amount of ejecta the surface produces. The value of  $\mathcal{C}$ used here characterizes impacts of glass projectiles into ice-silicate surfaces. The ejecta mass production rate, $\mathcal{P}^+$, is likely orders of magnitude smaller for regolith dominated surface than  those covered by water-ice surfaces, as suggested by the discrepancy of the meteoroid modeling and Lunar Dust Experiment discussed in \citet{Pokorny_etal_2019}. \added{See Section \ref{SEC:Discussion_Impacts} for a discussion on the scaling of impact processes for different surface types with respect to other methods and lab experiments.}}

While the area of craters produced by meteoroid bombardment $\mathcal{A}$ is a useful quantity, we also define the e-folding lifetime $T_\mathcal{A}$ that defines the time that craters produced by meteoroids  will take to cover $1-e^{-N}$ of the surface, where $N$ is the number of lifetimes the surface experienced. \added{Since $\mathcal{A}$ represents the area of craters per unit surface area pre time, it has units of s$^{-1}$, and then $T_\mathcal{A} = \mathcal{A}^{-1}$}.  This means that if $T_\mathcal{A}$ is 10 Myr, then in \added{10 Myr approximately 63\% of the surface will be covered by craters, while in }30 Myr $95\%$ of the surface will be covered by craters, \added{and so on}. 

In order to decrease the computation time we investigate only the centroids of each triangular surface element, thus $S(\lambda,\beta)$ is either one or zero. To evaluate this simplification we run our ray-tracing procedure using 10, 50, and 100 points per surface triangle. When we \replaced{averaged}{average} these results over all directions of the meteoroid environment, we saw only negligible changes in all the quantities analyzed here. This is in contrast to solar radiation simulations, whose source is a small disc ($<0.1^\circ$ in radius as viewed from Ceres) in the sky and requires finer resolution \citep{Mazarico_etal_2018}.


\subsection{Ceres' rotation, pole precession, obliquity oscillation}
The effect of Ceres' rotation period of $P_\mathrm{rot}=9.074$ hr is implemented by rotating the meteoroid map (i.e., the celestial sphere) along Ceres' rotational axis and averaging the contribution of different position into one mean map. Since our meteoroid model time stamps are recorded in 10 day intervals, our simplified rotation method introduces $\sim 2\%$ uncertainty/error to the daily flux, which is much smaller than the intrinsic uncertainties in the meteoroid model; for the model uncertainty discussion see \citet{Pokorny_etal_2018,Pokorny_etal_2019} and Section \ref{SEC:Discussion_MetModel}.

The precession of the pole is implemented by rotating the Ceres rotational axis around the axis perpendicular to the orbital plane and adopting a pole precession rate of 210 kyr \citep{Ermakov_etal_2017}. The default rotational axis pointing in our model is set to right ascension $\alpha = 291.42751^\circ$ and declination $\delta = 66.76043^\circ$, which translates to  ecliptic longitude $\lambda_\epsilon = 11.18622^\circ$ and latitude $\beta_\epsilon = 81.55038^\circ$, or longitude and latitude with respect to the orbital plane $\lambda_\mathcal{O}=328.24905^\circ$ and $\beta_\mathcal{O}=85.96854^\circ$. The rotation from ecliptic coordinates to orbital coordinates is $\mathcal{R}_z(-\omega)\mathcal{R}_x(-I)\mathcal{R}_z(-\Omega)$, where $\mathcal{R}_a(y)$ is the rotation matrix with respect to axis $a$ which represents a clock-wise rotation by an angle $y$. In a more specific way:

\begin{equation}
\begin{pmatrix}
\cos \lambda_\epsilon \cos \beta_\epsilon \\
\sin \lambda_\epsilon \cos \beta_\epsilon \\
\sin\beta_\epsilon
\end{pmatrix}
=
\begin{pmatrix}
   \cos\Omega & -\sin\Omega & 0 \\ 
\sin\Omega & \cos\Omega & 0 \\
0 & 0 & 1 
\end{pmatrix}
\begin{pmatrix}
 1 & 0 & 0 \\
 0 & \cos i & -\sin i \\
 0 & \sin i & \cos i
\end{pmatrix} 
\begin{pmatrix}
   \cos\omega & -\sin\omega & 0 \\ 
\sin\omega & \cos\omega & 0 \\
0 & 0 & 1 
\end{pmatrix}
\begin{pmatrix}
\cos\lambda_\mathcal{O} \cos \beta_\mathcal{O}\\
\sin\lambda_\mathcal{O} \cos \beta_\mathcal{O}\\
\sin\beta_\mathcal{O}
\end{pmatrix}
\end{equation}

The effects of obliquity/axial tilt are calculated for each obliquity value separately and then the contribution of each obliquity regime is weighted by the time Ceres spends in that obliquity regime, based on Fig. 1 in  \citet{Ermakov_etal_2017}.

\section{Results - Precession cycle average}\label{Sec:Preccesion_Cycle}
First, we look at results of the meteoroid bombardment of Ceres, averaged over the entire precession cycle, i.e., approximately 210 kyr \citep{Ermakov_etal_2017}. Here, we assume the proper orbital elements of Ceres\footnote{We use synthetic values from \url{ https://newton.spacedys.com/astdys/index.php?pc=1.1.6&n=Ceres}}, with $a=2.7671$ au, $e=0.116198$, and $\sin(i)=0.167585$ ($i=9.64744^\circ$).

\begin{figure}
\epsscale{1.1}
\plotone{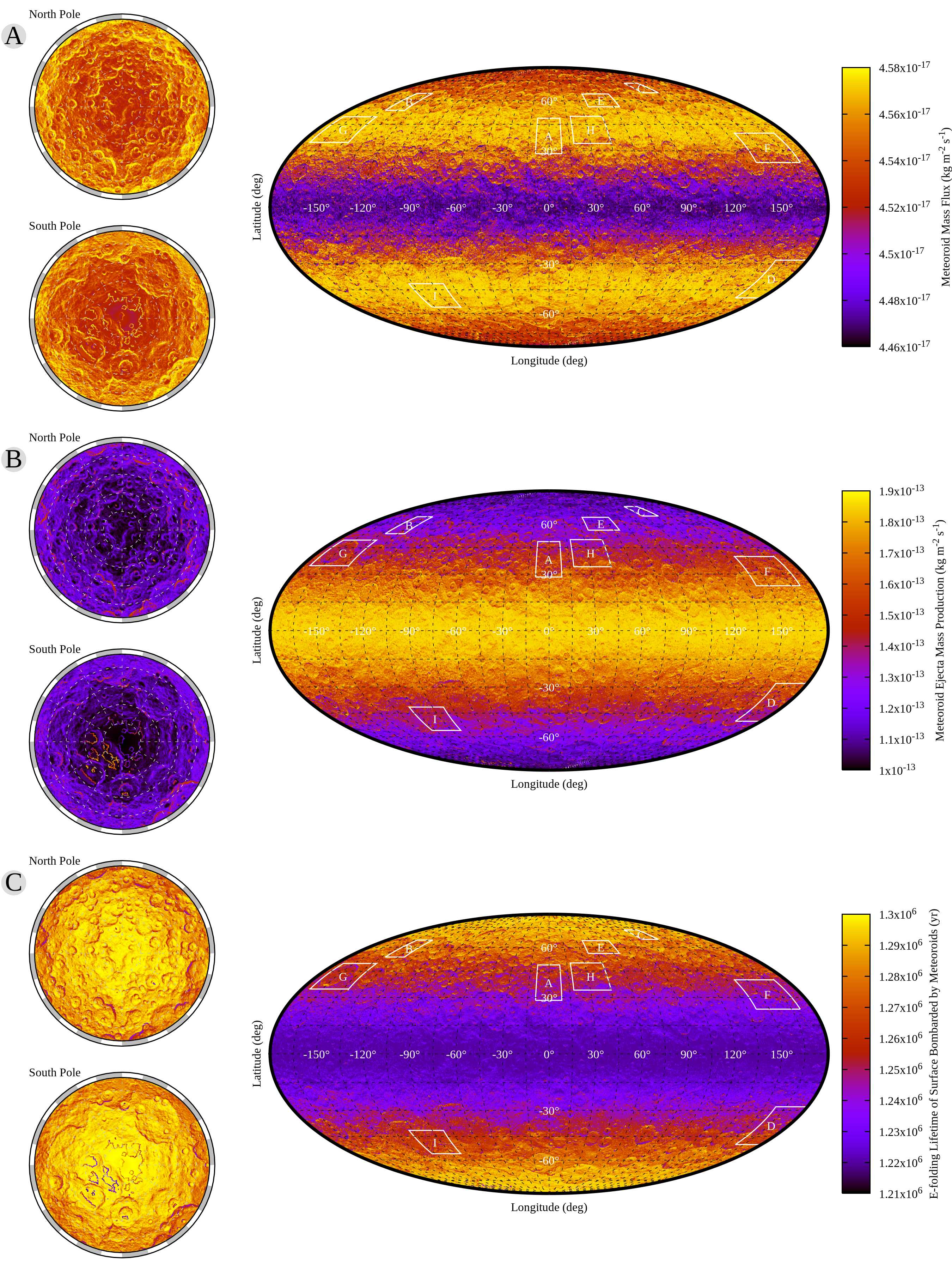}
\caption{
\label{FIG:Map_Averages}
(A): Global map of the meteoroid mass flux $\mathcal{M}$ on Ceres with nine areas of interest from \citet{Combe_etal_2019} highlighted (white areas with white labels). The left side of the figure shows the north/south pole views in Cartesian coordinates showing areas within $30^\circ$ away from the pole. The meteoroid environment is averaged over the entire precession cycle.
(B): The same but for ejecta mass production rate $\mathcal{P}^+$ on Ceres.
(C): The same but for surface e-folding lifetime on Ceres. All units are in SI except for the surface e-folding lifetime which we show in years.
}
\end{figure}

Figure \ref{FIG:Map_Averages}A shows in detail the entire surface of Ceres color coded by the value of the meteoroid mass flux $\mathcal{M}$. Despite\deleted{,} the seemingly significant difference between the equatorial and polar regions, the average $\mathcal{M}$ is almost constant over the entire surface, where the difference between the maximum and minimum values is 7.8\%. The average value over the entire surface is $\overline{\mathcal{M}}=4.54\pm0.04\times
10^{-17}$ \KGMS, where $\mathcal{M}$ peaks at mid-latitudes, around $60^\circ$. The local topography of Ceres plays a negligible role when the meteoroid environment is averaged over a longer timescale. This is also apparent on both Cererean poles (right side in Fig. \ref{FIG:Map_Averages}A), where we see $<2\%$ differences between maximum and minimum values.


We also highlight nine areas of interest from \citet{Combe_etal_2019} that showed detections of exposed \HHO from Visible and InfraRed (VIR, the mapping spectrometer of the Dawn mission) remote sensing observations (white labeled rectangles in Fig. \ref{FIG:Map_Averages}). The coordinates of these nine areas are shown in Table \ref{TABLE:Area_Description}. 
Unlike solar irradiation which is almost a point source at the distance of Ceres, the meteoroid environment is able to influence any surface element on the Cererean surface and even the most significant features do not provide permanent shadowing from the meteoroid bombardment. \citet{Ermakov_etal_2017} identified seven bright crater floor deposits (BCFDs) that are correlated with the most persistend PSRs. Since all of these seven areas have high latitudes ($|\beta|>69.7^\circ$), they are all comparable to area (C) - Messor crater from \citet{Combe_etal_2019}.

The global map of the  ejecta mass production rate $\mathcal{P}^+$ is shown in Figure \ref{FIG:Map_Averages}B. For $\mathcal{P}^+$ the equatorial areas experience the maximum exposure, while the polar regions experience 43\% less exposure (i.e., a factor 1.76 difference). A similar situation holds for the meteoroid energy flux, where the pole receives $\sim19\%$ less $\mathcal{E}$ than the equatorial regions. The reason for the difference between the global behavior of $\mathcal{M}$ and $\mathcal{P}^+$ shown Fig. \ref{FIG:Map_Averages}A is due to the different impact velocities of meteoroid populations. As shown in Fig. \ref{FIG:FLUX_vs_Orbit} the HTC and OCC meteoroids have the highest velocities with median velocities 2.1-2.6 times larger than the JFC population that dominates the meteoroid mass flux. High velocity meteoroids from both HTC and OCC populations are impacting Ceres preferentially from the invariable plane which is close to the ecliptic. Due to their high impact velocities, these meteoroids are not sensitive to Ceres' own orbital velocity variations. This results in a continuous high energy bombardment of the equatorial regions, while polar regions experience high energy impacts on grazing angles. The mean values for the meteoroid energy flux and ejecta production rate are $\overline{\mathcal{E}}=5.16\pm0.31\times
10^{-16}$ MJ s$^{-1}$ m$^{-2}$ and $\overline{\mathcal{P}^+}=1.56\pm0.22\times
10^{-16}$ kg s$^{-1}$ m$^{-2}$, respectively.

The last quantity we discuss here is the surface e-folding lifetime $T_\mathcal{A}$ (Fig. \ref{FIG:Map_Averages}C). The shortest e-folding times are in the equatorial area, while $T_\mathcal{A}$ increases toward the both poles. However, the difference between the maximum and minimum values of $\mathrm{A}$ is small: $<6\%$. Unlike $\mathcal{P}^+$ and $\mathcal{E}$ the e-folding time is not modulated by the impactor velocity, since we assume a constant impactor-to-crater area ratio $F_\mathrm{cr}=63$ (see Sec. \ref{SEC:Definitions}). The mean e-folding time $\overline{\mathcal{T}_\mathcal{A}}=1.25 \pm0.02\times 10^{6}$ yr means that in 3.75 Myr 95\% of the surface should be covered by meteoroid-induced craters. This does not include impacts of secondary ejecta \citep[secondaries; see e.g.,][]{Costello_etal_2018, Costello_etal_2020}. The typical depth of meteoroid-induced craters is a factor \replaced{of a few larger}{2--4} larger than the particle radius \citep{Koschny_Grun_2001}. When we average over the entire size range simulated here, we find a mean crater depth around $100-300~\mu$m. The mean crater depth depends on the location of Ceres as well as the meteoroid population causing the impacts. This is due to the different size-frequency distributions that meteoroid populations simulated here have when impacting Ceres. Note, that the \citet{Koschny_Grun_2001} explored only impact velocities up to 10 \KMS{}, so the crater depth has high uncertainty that we cannot quantify at the moment.

A closer look at the nine areas showing exposed \HHO signatures, the crater morphology and the general topography have a rather small effect even for the Juling crater that shows the largest ratio between the maximum and minimum values for all three quantities: $\mathcal{R}_\mathcal{M}=\mathcal{M}_\mathrm{max}/\mathcal{M}_\mathrm{min}=1.03$, $\mathcal{R}_\mathcal{E} = 1.24$, $\mathcal{R}_\mathcal{P}^+=1.75$ and $\mathcal{R}_{\mathcal{T}_\mathcal{A}}=1.06$. The only exception is the ejecta mass production rate $\mathcal{P}^+$, which for the Juling crater shows 75\% difference between the crater rim $\mathcal{P}^+ = 1856\times 10^{-16}$ \KGMS{} and partially shadowed crater floor $\mathcal{P}^+ = 1063\times 10^{-16}$. This is due to its cubic dependence on the cosine of the incident angle and the $v_\mathrm{imp}^{2.46}$ velocity scaling, which emphasize the effect of fast meteoroids close to the ecliptic. The surface features facing the ecliptic produce significantly more ejecta than those that are effectively shadowed by the crater rim. This effect is most efficient for mid-latitudes $30^\circ<|\beta|<50^\circ$, where it provides $\sim70\%$ difference between the exposed and shadowed portion of the crater. This efficiency of shadowing drops toward the equator and both poles, where for the Messor crater we see a drop to $49\%$ difference between the exposed/shadowed areas. 
\begin{deluxetable}{lcclcccc}

\tablecaption{\label{TABLE:Area_Description} Description of \citet{Combe_etal_2019} areas with their identifier used in this article, their surface longitude and latitude in degrees, feature name and four quantities defined in Sec. \ref{SEC:Definitions}. }

\tablenum{2}

\tablehead{\colhead{Area} & \colhead{Longitude} & \colhead{Latitude} & \colhead{Feature name} & 
\colhead{$\mathcal{M}$ $\times 10^{-16}$ } & 
\colhead{$\mathcal{E}$ $\times 10^{-16}$ } & 
\colhead{$\mathcal{P^+}$ $\times 10^{-16}$ } & 
\colhead{$\mathcal{T}_\mathcal{A}$} \\
 ID & (deg) & (deg) & & kg m$^{-2}$ s$^{-1}$ & MJ m$^{-2}$ s$^{-1}$ & kg m$^{-2}$ s$^{-1}$ & Myr
} 

\startdata
A & -0.30  &  38.60& Oxo crater     & 0.457 &      50.70 &    1492.33 & 1.252 \\  
B &-138.05 & 59.40&   ---           & 0.456 &      47.50 &    1263.96 & 1.276 \\
C &113.89  &69.25& Messor crater    & 0.454 &      46.12 &    1165.53 & 1.287 \\
D &168.80 &-39.00  & Juling crater  & 0.457 &      50.55 &    1481.66 & 1.253 \\
E & 50.80  &60.35  &    ---         & 0.456 &      47.31 &    1250.77 & 1.277 \\
F &155.95  &31.80  &    Inkosazana crater         & 0.456 &      52.02 &    1589.75 & 1.242 \\
G &-160.05  &42.15 & Ezinu crater   & 0.457 &      49.95 &    1437.68 & 1.257 \\
H & 31.45  &42.10 &     ---         & 0.457 &      50.39 &    1470.66 & 1.254 \\
I &-94.75 &-48.55  & Baltay crater  & 0.457 &      49.21 &    1385.75 & 1.263
\enddata




\end{deluxetable}










\section{Results - One orbit of Ceres}\label{SEC:One_Period}
The image of the meteoroid environment imprint on Cererean surface averaged over one precession cycle of Ceres greatly simplifies the picture. The averaged picture is important for understanding the long-term surface evolution, but Ceres experiences significant meteoroid flux changes during the Cererean year (Fig. \ref{FIG:FLUX_vs_Orbit}). In this Section we analyze variations of several quantities over one orbit of Ceres, 4.61 years. Such a time segment is longer than the duration of the scientific stay of the Dawn spacecraft at Ceres (RC3 orbit phase started on April 23, 2015 and on October 31, 2018 the spacecraft ran out of its propellant). 

\begin{figure}
\epsscale{1.1}
\plotone{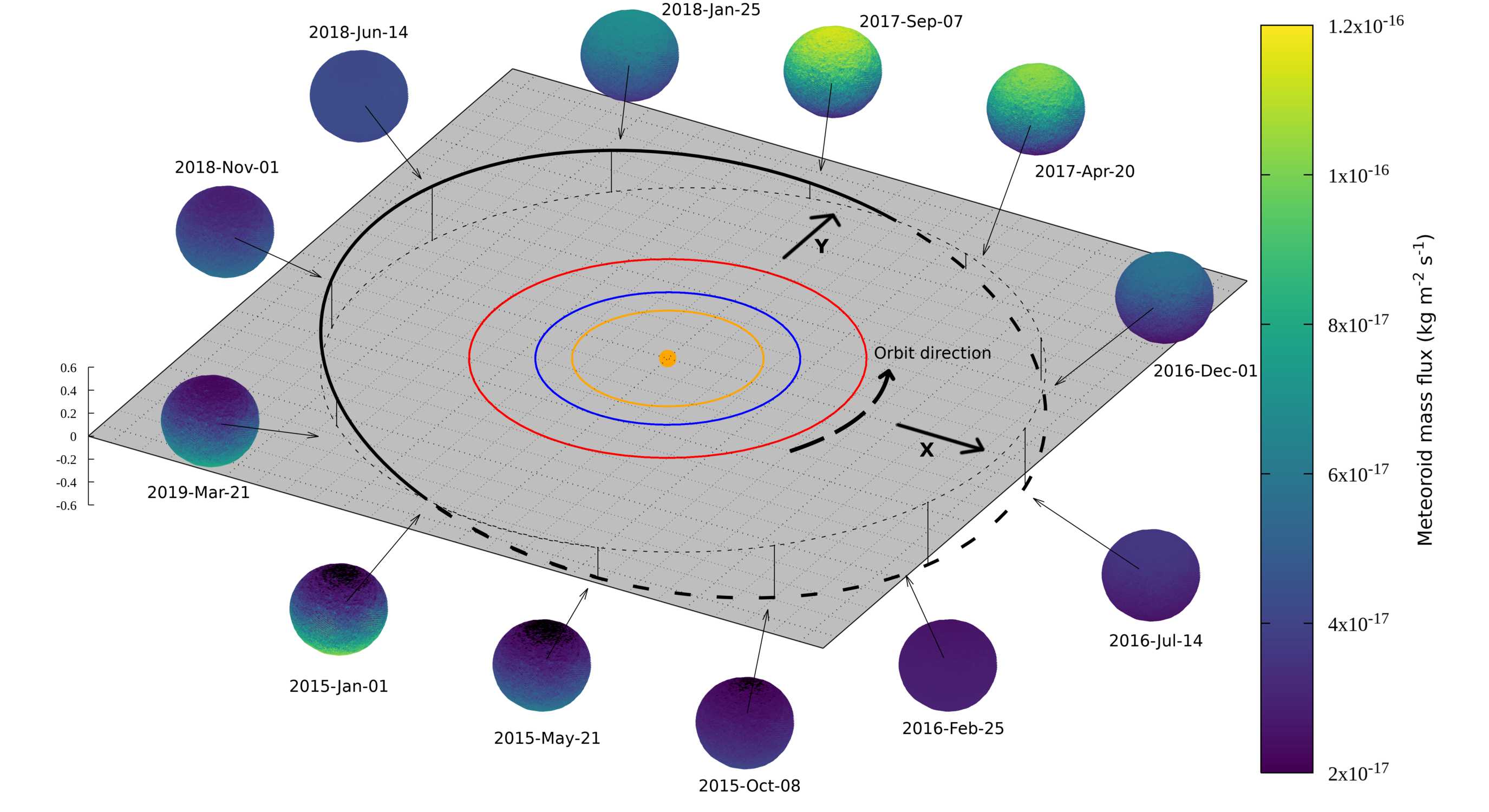}
\caption{
\label{FIG:Orbit_Diagram}
Orbital diagram of Ceres showing 12 different instances (time stamps) of its orbit around the Sun. The colored circles around the origin show the orbits of Venus (orange), Earth (blue), and Mars (red). The orbit of Ceres is represented by a black solid line when the asteroid is above the ecliptic, and by the black solid dashed line when below the ecliptic. Ceres' surface is color-coded by the value of meteoroid mass flux in kg per meter square per second. 
}
\end{figure}

In Fig. \ref{FIG:FLUX_vs_Orbit} we \replaced{showed}{show} that the meteoroid mass flux $\mathcal{M}$ impacting Ceres is most significantly modulated by the asteroid's distance from the ecliptic. To put this in a different perspective, we show the orbit of Ceres in Cartesian coordinates in Fig. \ref{FIG:Orbit_Diagram}. This Figure shows 12 different time records of $\mathcal{M}$ on Cererean surface averaged over one rotation period (9.1 hours) in SI units from January 1st, 2015 to March 21st, 2019. In Fig. \ref{FIG:Orbit_Diagram} we see the effects of the orbital motion of Ceres on the global shape of the meteoroid mass flux. On January 1st, 2015 Ceres just passed the descending node and its negative $z$-axis velocity is close to its maximum, $v_z=-3.32$ \KMS (note that the median impact velocity of JFC meteoroids is $\overline{V_\mathrm{50\%}}\mathrm{(JFC)}=9.32\pm0.46$ \KMS). This increases the number of impacts to the southern hemisphere, since Ceres is plunging downwards and the relative velocity of meteoroids impacting the southern hemisphere is increased, while the northern hemisphere experiences attenuated meteoroid mass flux. The opposite effect can be seen on April 20th, 2017, when Ceres is close to the ascending node and its positive $z$-axis velocity is close to its maximum $v_z=3.40$ \KMS. When the $z$-axis velocity is close to zero, i.e., the sum of the argument of perihelion and true anomaly is $\sin(\omega+f)=0$, the meteoroid mass flux is symmetric around the Cererean equator as seen on July 14th, 2016 and June 14th, 2018. 

\begin{figure}
\epsscale{1}
\plotone{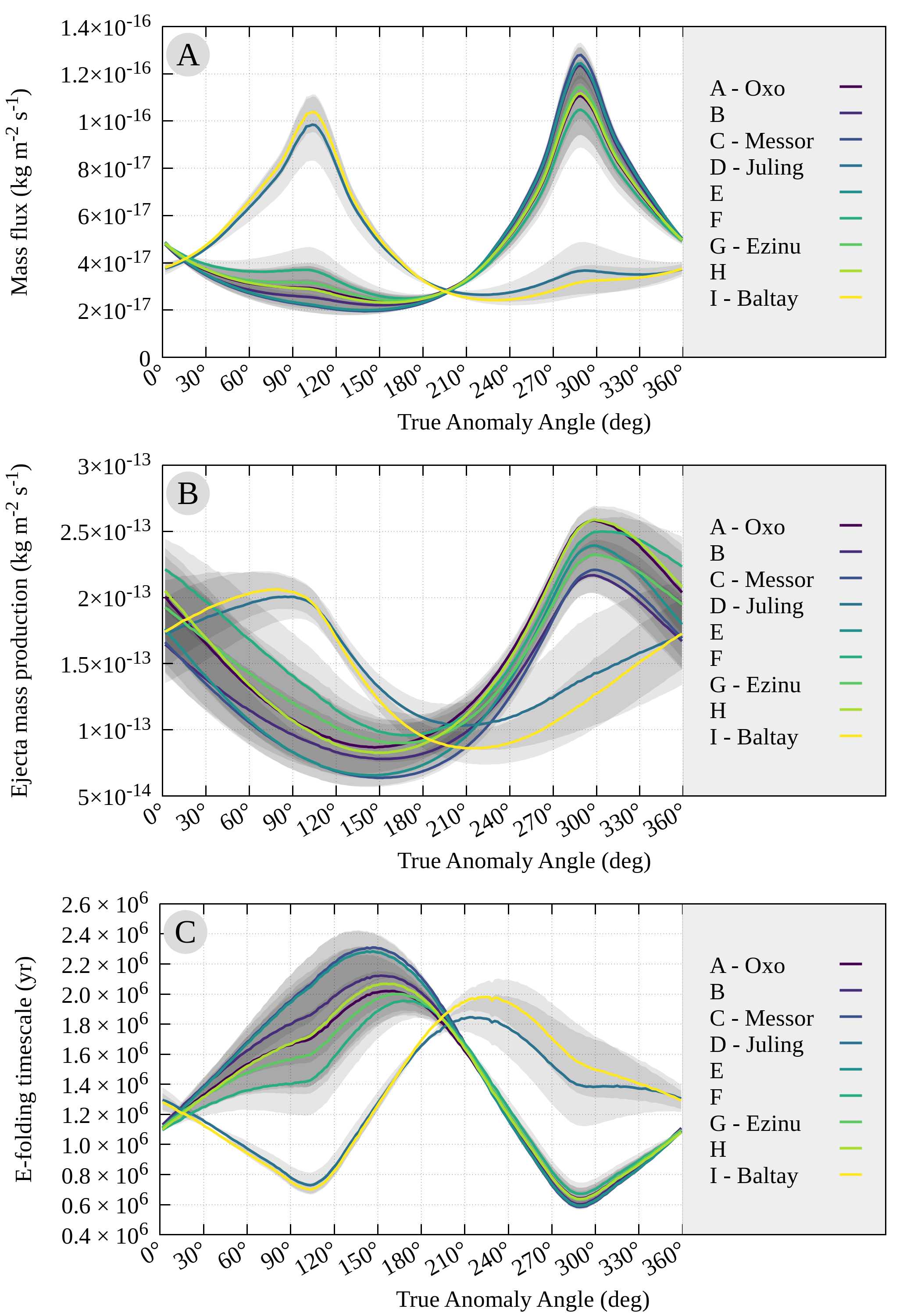}
\caption{
\label{FIG:9area_line}
Top: Variations of the meteoroid mass flux $\mathcal{M}$ with Ceres' true anomaly angle for nine areas showing signatures of \HHO based on \citet{Combe_etal_2019} (see Table \ref{TABLE:Area_Description} for description). The two peaks in $\mathcal{M}$ correspond to the ecliptic crossings and the maximum flux from MBA and JFC meteoroid populations. 
Middle: The same as the top panel but now for the ejecta mass production rate $\mathcal{P}^+$. 
Bottom: The same as the top panel but now for the surface exposure e-folding time $\mathcal{T}_\mathcal{A}$. 
}
\end{figure}

The variations of $\mathcal{M}$, $\mathcal{P}^+$, and $\mathcal{T}_\mathcal{A}$ for the \citet{Combe_etal_2019} areas (Table \ref{TABLE:Area_Description}) are shown in Fig. \ref{FIG:9area_line}. The mass flux during one orbit of Ceres shows a double peaked structure, where depending on the latitude of the feature, the global maximum occurs at true anomaly angle TAA = $106^\circ$ (for southern hemisphere features D - Juling and I - Baltay) or at TAA = $286^\circ$ (for nothern hemisphere features) coinciding with the ecliptic crossing (Fig. \ref{FIG:9area_line}A). These mass flux spikes are correlated with the density of the zodiacal cloud that is densest close to the ecliptic and gets more tenuous further from the ecliptic. The north/south hemisphere seasonality is caused by the orbital velocity in the $z$-axis described in Fig. \ref{FIG:Orbit_Diagram}. As shown in Fig. \ref{FIG:FLUX_vs_Orbit}, the TAA = $286^\circ$ peak is stronger than the TAA = $106^\circ$ peak due to Ceres' smaller heliocentric distance; the zodiacal cloud gets denser with decreasing heliocentric distance. The north/south asymmetry is changing as Ceres undergoes nodal precession and when averaged over the entire precession cycle the difference between the areas in the south and north are negligible (see Table \ref{TABLE:Area_Description}).

The ejecta mass production rate $\mathcal{P}^+$ true anomaly profile differs from that of $\mathcal{M}$ mainly due to two factors. First, $\mathcal{P}^+$ is strongly dependent on the impact velocity of individual meteoroids ($v_\mathrm{imp}^{2.46}$, see Eq. \ref{EQ:EJECTA}). The highest impact velocities come from OCC and HTC meteoroids (see median velocities for each population in Sec. \ref{SEC:Meteoroid_Model}), which shifts the dominance over $\mathcal{P}^+$ from JFCs and MBAs to long-period comet particles. As we \replaced{showed}{show} in Fig. \ref{FIG:FLUX_vs_Orbit}, the long-period comets are not very sensitive to the distance from the ecliptic which attenuates the variations of $\mathcal{P}^+$ during its orbit to a factor of 2-3 (compared to a factor of 4-7 variations of $\mathcal{M}$). Second, the 5-95 percentile (gray shaded area in Fig \ref{FIG:9area_line}) is wider than for $\mathcal{M}$ due to the $\cos^3$ of incidence angle dependence. This accentuates the effect of surface features, where surface patches experiencing almost perpendicular impacts  produce significantly more ejecta than those subjected to grazing impacts. 

The surface e-folding timescales $\mathcal{T}_\mathcal{A}$ show a similar trend in the north/south asymmetry as the two previous values, but their maxima are inverse with respect to the true anomaly angle (Fig. \ref{FIG:9area_line}C). This is understandable, since the higher impactor flux produces shorter e-folding timescales. $\mathcal{T}_\mathcal{A}$ is not sensitive to the impact velocity, thus is similar to the mass flux variations with slightly smaller minimum/maximum ratios of 3-4.

\begin{figure}
\epsscale{1}
\plotone{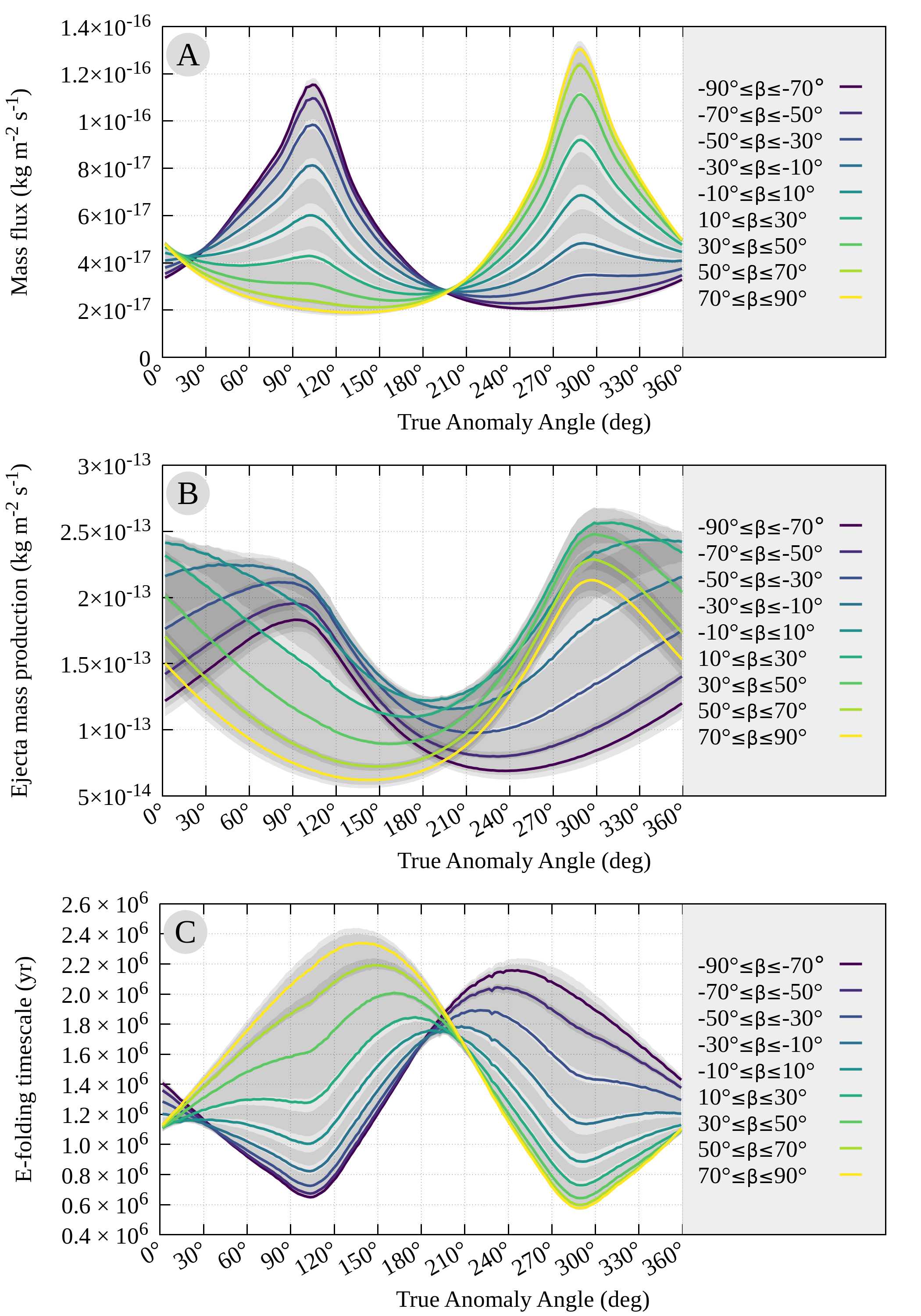}
\caption{
\label{FIG:Latitude_line}
Top: Variations of the meteoroid mass flux $\mathcal{M}$ with Ceres' true anomaly angle for nine latitudinal stripes with $20^\circ$ width. The two peaks in $\mathcal{M}$ correspond to the ecliptic crossings and the maximum flux from MBA and JFC meteoroid populations. 
Middle: The same as the top panel but now for the ejecta mass production rate $\mathcal{P}^+$. 
Bottom: The same as the top panel but now for the surface exposure e-folding time $\mathcal{T}_\mathcal{A}$. 
}
\end{figure}

Analysis of the \citet{Combe_etal_2019} areas showed significant variations of all quantities analyzed here during one Ceres' orbit. From Fig. \ref{FIG:9area_line} we can infer that areas on similar latitudes are undergoing very similar variations within one orbit. Since the \citet{Combe_etal_2019} areas are preferentially closer to the Cererean poles, many potentially interesting areas are left out. Furthermore, due to the short rotation period of Ceres, that longitudinally averages meteoroid bombardment effects at a given latitude, we can simply divide Ceres into latitudinal strips and analyze the entire dwarf planet as a whole. In Figure \ref{FIG:Latitude_line} we show variations of $\mathcal{M}$, $\mathcal{P}^+$, and $\mathcal{T}_\mathcal{A}$ with true anomaly angle for $20^\circ$-wide latitudinal strips, revealing the variations experienced by the entire surface. Lines in Fig. \ref{FIG:Latitude_line} can be used to determine $\mathcal{M}$, $\mathcal{P}^+$, and $\mathcal{T}_\mathcal{A}$ for any point on the surface. The uncertainty of this approximation is quite small, because for each latitudinal strip the difference between the median value (color coded lines in Fig. \ref{FIG:Latitude_line}) and the 5\% or 95\% percentile is $<35\%$.  

Figure \ref{FIG:Latitude_line} shows that the equatorial latitudes experience smaller variations in all quantities analyzed here, where the ratio between the maximum and minimum value during one orbit increases with latitudes closer to the Cererean poles. For example, the $\mathcal{M}$ max-min ratio for $-10^\circ<\beta<10^\circ$ is $2.43$, while the same quantity for the north pole areas $70^\circ<\beta<90^\circ$ is $6.88$. As mentioned before, the north pole peak values are smaller than those on the south pole because Ceres passes through the ascending node closer to the Sun, hence, through the denser portion of the Zodiacal cloud.

Telescopic observations of the water exosphere on Ceres by \citet{Ahearn_Feldman_1992, Kuppers_etal_2014} suggest production rates of $3-6$ kg s$^{-1}$. Let us assume an extreme case where all ejecta produced by meteoroids convert to the detectable water exosphere. We can obtain the water exosphere production rate via multiplying $\mathcal{P}^+$ by the surface area covered with surface water-ice. Using our values of $\mathcal{P}^+=0.5-2.5 \times 10^{-13}$ \KGMS, this would require a surface water-ice area of $1.2-12 \times 10^{13}$ m$^{2}$ to obtain production rates $3-6$ kg s$^{-1}$, which is equivalent to a sphere with a radius $1000-3000$ km. This would mean that even if the entire surface of Ceres was covered by water-ice, the meteoroid impacts would not be able to produce enough ejecta required to sustain the water exosphere observed by \citet{Ahearn_Feldman_1992, Kuppers_etal_2014}. 
Suppose that the entire Occator crater with diameter of 92 km is covered in water-ice and continuously bombarded by meteoroids. The Occator crater area is about $6.65\times 10^{9}$ m$^{2}$ and the ejecta mass produced per second is then $0.3-1.7\times 10^{-3}$ kg s$^{-1}$. Such a hypothetical value is much smaller than the sublimation from known water-ice patches \citep[0.16 kg s$^{-1}$][]{Landis_etal_2019}.

In the next Section we will compare these values with the identical quantities on Mercury and Moon and draw conclusions for the meteoroid bombardment effects on these three bodies.

\section{Comparison of Ceres to Mercury and Moon}\label{SEC:Comparison}
Similarly to Sec. \ref{Sec:Preccesion_Cycle}, we \replaced{calculated}{calculate} the effects of meteoroid bombardment using our meteoroid model on surfaces of Mercury and the Moon \citep{Pokorny_etal_2018,Pokorny_etal_2019,Pokorny_etal_2020}. For Mercury we \replaced{used}{use} the following orbital elements and escape velocity: $a =0.3871$ au, $e=0.2056$, $i=7.0056^\circ$, and $V_\mathrm{esc}=4.25$ \KMS, whereas for the Moon we \replaced{used}{use} $a=1.0000$ au,	$e=0.0167$,	$i=0.0005^\circ$, and $V_\mathrm{esc}=2.38$ \KMS. The remaining orbital elements \replaced{were}{are} assumed to be randomly distributed between $0$ and $360^\circ$. In order to simplify our calculation, we \replaced{used}{use} the shape model of Ceres as a substitute for the shape models of Mercury and Mars. Since in this Section we aim to analyze the global models of three airless bodies, we assume that simplification is acceptable. To further check our assumptions, we \replaced{compared}{compare} our results to those calculated on a smooth sphere representing Mercury and the Moon. The values shown in this Section agree within 1\%. Note that all three objects rotate much faster than their nodal precession periods and they are not in 1:1 spin-orbital resonance with the Sun and we can approximate that the meteoroid bombardment is longitudinally uniform. This ensures that the surface of each of these three bodies is uniformly affected with respect to the surface longitude. The 3:2 spin-orbit resonance of Mercury with respect to the Sun can possibly introduce some secondary non-uniform variations in effects of meteoroid bombardment. Mercury is still rotating with respect to the meteoroid environment, thus we expect that the effects of the meteoroid bombardment average out. We reserve the quantification of the 3:2 spin-orbit resonance for future work due to the necessity of simulating the entire Hermean two-year cycle in fine detail.

\begin{figure}
\epsscale{1}
\plotone{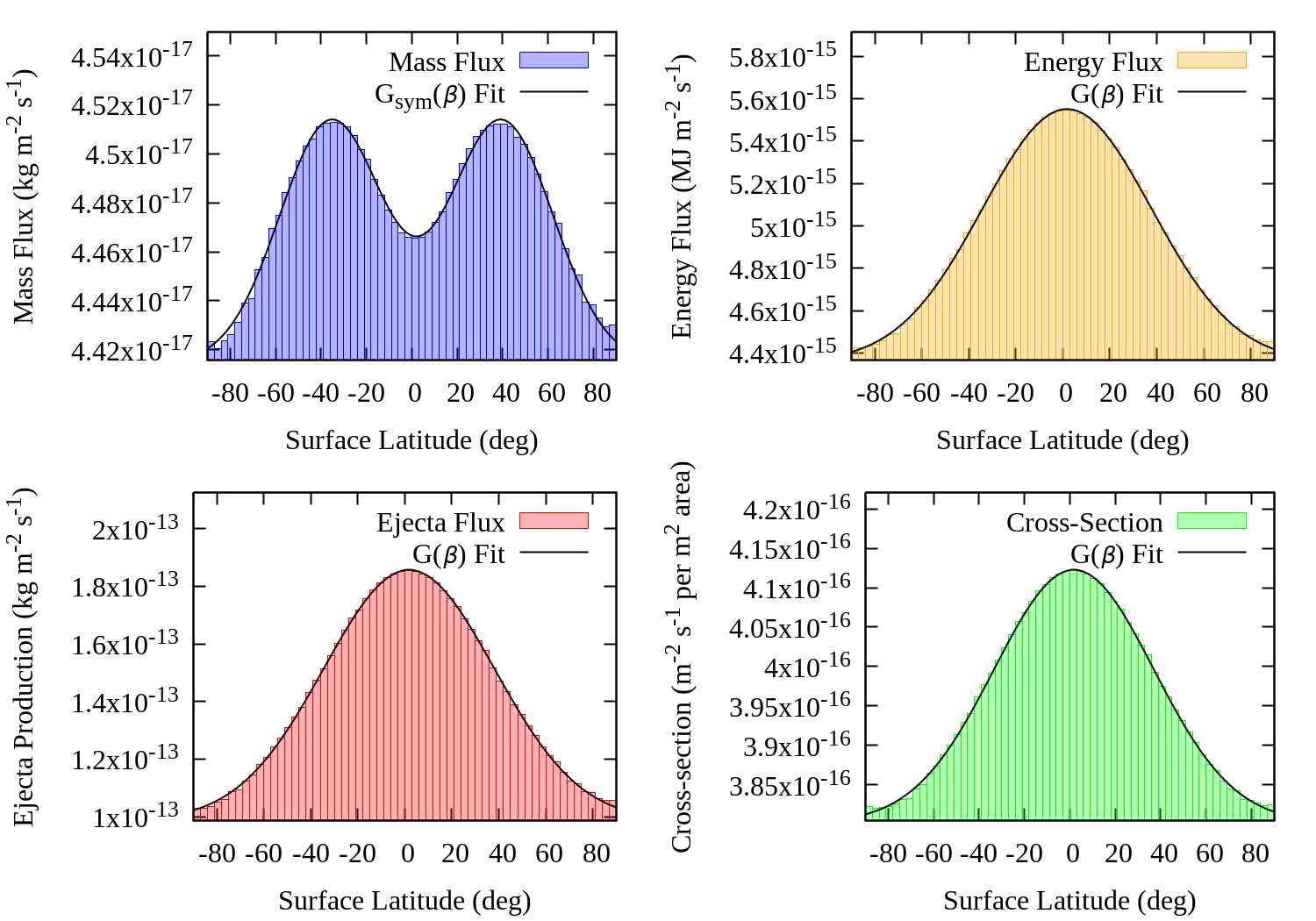}
\caption{
\label{FIG:Ceres_Latitude_FITS}
Variations of the meteoroid mass flux $\mathcal{M}$ (top left), meteoroid energy flux $\mathcal{E}$ (top right), ejecta mass production rate $\mathcal{P}^+$ (bottom left), and the area of craters produced by meteoroid impacts $\mathcal{A}$ (bottom right) with the latitude on Ceres for 60 latitudinal strips. Functions of $\mathcal{M}$, $\mathcal{E}$, and $\mathcal{P}^+$ are fitted using Eq. \ref{EQ:Gaussian}, while $\mathcal{A}$ is fitted using Eq. \ref{EQ:Cosine}. Fits for all quantities are represented by solid black lines.}
\end{figure}

For each of the three airless bodies, we \replaced{analysed}{analyze} 60 segments that \replaced{were}{are} uniformly distributed in surface latitude $\beta$ (i.e, $3^\circ$ wide segments). The results for Ceres are in Fig. \ref{FIG:Ceres_Latitude_FITS}, where we show variations of $\mathcal{M}$, $\mathcal{E}$, $\mathcal{P}^+$, and $\mathcal{A}$ with respect to surface latitude. To make our results more readable and transferable, we fit simple four-parameter functions to each quantity. For $\mathcal{M}$, $\mathcal{E}$, $\mathcal{P}^+$ we \replaced{achieved}{achieve} a best fit using the Gaussian function $G(\beta)$:

\begin{equation}
    G(\beta)= a_0 \exp \left[-0.5\left(\frac{\beta-\mu_0}{\sigma_0}\right)^2\right]+b_0,
    \label{EQ:Gaussian}
\end{equation}

\noindent where $a_0$ is the maximum amplitude of the function occurring at $\beta = \mu_0$, $b_0$ is the offset of the function in the $y$ direction, $\mu_0$ is the offset from the center in the $x$-axis (i.e., in the latitude), and $\sigma_0$ is the standard deviation. From Figure \ref{FIG:Ceres_Latitude_FITS}, it is evident that $\mathcal{M}$ is not a Gaussian distribution, but rather resembles the sum of two symmetric Gaussians. For fitting such a profile, we use a four-parameter function
$G_\mathrm{sym}(\beta)$:

\begin{equation}
    G_\mathrm{sym}(\beta)= a_0 \exp \left[-0.5\left(\frac{\beta-\mu_0}{\sigma_0}\right)^2\right]+a_0 \exp \left[-0.5\left(\frac{\beta+\mu_0}{\sigma_0}\right)^2\right]+b_0,
    \label{EQ:Gaussian_Symmetric}
\end{equation}

\noindent where the only difference between the two Gaussians is the sign at $\mu_0$. Furthermore, $\mathcal{A}$ on Mercury \replaced{responded}{responds} the best to fitting a cosine function
$C(\beta)$:

\begin{equation}
    C(\beta)= a_0 \cos \left(\frac{\beta-\mu_0}{\sigma_0}\right)+b_0,
    \label{EQ:Cosine}
\end{equation}

\noindent where $a_0$ is the amplitude, $b_0$ is the offset of the function in the $y$ direction, $\mu_0$ is the offset from the center in the $x$-axis (i.e., in the latitude), and $\sigma_0$ describes the period of the cosine. These three functions \replaced{were}{are} chosen for their simplicity and ease of interpretation. We tested more than 80 other probability density distributions available in the \texttt{SciPy} framework, but we \replaced{found}{find} no distribution that would provide significantly better fits than our three functions.


Table \ref{TABLE:Comparison_Mer_Moon_Ceres} shows the results of our function fitting to the latitudinal distributions of $\mathcal{M}$, $\mathcal{E}$, $\mathcal{P}^+$, and $\mathcal{A}$ for Mercury, Moon, and Ceres. For each planetary body, we show the four fit parameters, the difference between the maximum and the minimum value, and the ratios, $R$, between the three objects for the average values of quantities $\mathcal{M}$, $\mathcal{E}$, $\mathcal{P}^+$, and $\mathcal{A}$. Mercury and Moon both have $\mu_0 = 0.0$ except for cases when we fit $G_\mathrm{sym}$. This is because we \replaced{enforced}{enforce} the latitudinal symmetry in our fits since both bodies have negligible flatness (i.e., almost perfectly spherical shape).

Mercury experiences the highest values of all quantities analyzed here, which are an order of magnitude higher than those at the Moon, and 2-3 orders of magnitude higher than those at Ceres. The meteoroid energy flux $\mathcal{E}$ and the ejecta mass production rate $\mathcal{P}^+$ at Mercury are amplified  with respect to those at Ceres due to a combination of higher impact velocities and meteoroid fluxes closer to the Sun. Mercury, assuming that both Mercury and Ceres have the same surface material, produces $\sim 700$ times more ejecta than Ceres via meteoroid impacts. Similarly, the $\mathcal{E}$ is $\sim 400$ times higher on Mercury with respect to Ceres. These quantities have been shown to have strong correlation with the existence of a tenuous dust cloud around the Moon \citep{Horanyi_etal_2015, Pokorny_etal_2019, Szalay_etal_2015}, sustaining an exosphere of several metals around Mercury \citep{Killen_Hahn_2015,Merkel_etal_2017,Pokorny_etal_2018}, and potential stability of water-ice in permanently shadowed regions at Mercury and the Moon \citep{Pokorny_etal_2020, HAYNE_ETAL_2015,Deutsch_etal_2019}. 

The lunar surface water-ice stability was shown to suffer similar erosion rates from meteoroid bombardment \citep[about $13 \times 10^{-8}$ m per year;][]{Pokorny_etal_2020} as those from H Ly-$\alpha$ radiation \citep[about $7 \times 10^{-11}$ m per year;][]{Morgan_Shemansky_1991}. This means that the surface water-ice on Ceres should be primarily excavated by extrasolar radiation because the meteoroid energy flux is $\sim 25$ times smaller on Ceres as compared to that on the Moon. Meanwhile, the meteoroid bombardment should dominate the shadowed surface water-ice excavation on Mercury, due to an order of magnitude higher value of $\mathcal{E}$ as compared to effects of H Ly-$\alpha$. The quantification of the surface water-ice loss through meteoroid impacts at Ceres is a complex process that involves the combination of impact vaporization \citep[e.g., Eq. 10 in ][]{Cintala_1992} and loss of ejecta produced in impacts. Since both of these effects depend linearly on the meteoroid mass flux, scale with some power of impact velocity, and depend on material characteristics, it is beyond the scope of this article to obtain more concrete values. We can only conclude that the loss of surface water-ice through meteoroid impacts at Ceres is an order of magnitude smaller compared to that on the Moon, and 2-3 orders of magnitude smaller with respect to that at Mercury.

The area of craters produced by meteoroid bombardment $\mathcal{A}$ shows that fresh deposits on the Hermean surface are covered by meteoroid-induced craters $74\times$ faster than those on Ceres, whereas the lunar surface shows $12\times$ faster rates. This means that while at Ceres the mean surface e-folding time is  $\overline{\mathcal{T}_\mathrm{A}}=1.25$ Myr, the mean value for the Moon is 107 kyr and for Mercury 17.1 kyr. These values ignore the effect of secondary impactors, which might enhance the gardening rates on the surface \citep[for the effect on the Moon see][]{Costello_etal_2018}. As such, 95\% of the surface of Mercury's surface is covered by meteoroid-induced craters within three e-folding lifetimes, i.e., in 50 kyr. However, the crater depth resulting from meteoroid bombardment in the size range in our model is $<1$ mm according to laboratory experiments in \citet{Koschny_Grun_2001}. One caveat of such an experiment is the absence of $V_\mathrm{imp}>10$ \KMS\ impacts that are dominating the impacts on lunar and hermean surfaces, thus the penetration depths could significantly change with new laboratory experiments.

\begin{deluxetable}{c|crrccr||ccc|c}

\tablecaption{\label{TABLE:Comparison_Mer_Moon_Ceres} Comparison of four different quantities: rows (1,5,9) meteoroid mass flux $\mathcal{M}$, rows (2,6,10) meteoroid energy flux $\mathcal{E}$, rows (3,7,11) ejecta mass production $\mathcal{P^+}$, and rows (4,8,12) meteoroid\added{-induced crater} cross-section $\mathcal{A}$ at Mercury, Moon, and Ceres. Column (1) represents the selected quantity $\mathcal{Q}$, columns (2-5) show the function fit parameters, column (6) shows the percent difference between the maximum and minimum values, and columns (7-9) show the ratios $R$ for all quantities in this table with respect to other airless bodies analyzed here. For instance, $R_\mathrm{Moon}$=5.45 in the first row indicates $\mathcal{M}_\mathrm{Mercu}/\mathcal{M}_\mathrm{Moon}$, the ratio between the average meteoroid mass flux on Mercury with respect to that on the Moon. The last column (10) shows the function $\mathcal{F}$ used to fit the quantity. Quantities $\mathcal{M}, \mathcal{E}$, and $\mathcal{P}^+$ are fitted using Gaussian distributions described in Eq. \ref{EQ:Gaussian}, whereas $\mathcal{A}$ is fitted using the cosine function described in Eq. \ref{EQ:Cosine}. Parameters $a_0$ and $b_0$ are in SI units, while $\sigma_0$ and $\mu_0$ are in degrees. }

\tablenum{3}

\tablehead{\colhead{} &
\colhead{$Q$} &
\colhead{$a_0$} &
\colhead{$b_0$} &
\colhead{$\sigma_0$} &
\colhead{$\mu_0$} &
\colhead{Max/Min} &
\colhead{$R_\mathrm{Mercu}$} &
\colhead{$R_\mathrm{Moon}$} &
\colhead{$R_\mathrm{Ceres}$} &
\colhead{$\mathcal{F}$}  }

\startdata
\multirow{4}{*}{\rotatebox[origin=c]{90}{Mercury}}
&$\mathcal{M}$ &	$-9.70^{-17} $&	$2.34^{-15} $&	21.94 &	0 &	4.1 \%&	1 &	5.45 &	51.7 & $G$ \\
&$\mathcal{E}$ &	$8.39^{-13} $&	$1.68^{-12} $&	40.05 &	0 &	41.7 \%&	1 &	17.9 &	432 & $G$\\
&$\mathcal{P^+}$ &	$9.32^{-11} $&	$5.77^{-11} $&	33.23 &	0 &	154.2 \%&	1 &	23.4 &	707 & $G$\\
&$\mathcal{A}$ &	$-1.30^{-15} $&	$2.91^{-14} $&	23.75 &	0 &	10.1 \%&	1 &	6.26 &	74 & $C$\\
\hline\multirow{4}{*}{\rotatebox[origin=c]{90}{Moon}}
&$\mathcal{M}$ &	$1.60^{-16} $&	$2.84^{-16} $&	99.34 &	0 &	12.2 \%&	0.184 &	1 &	9.48 & $G$\\
&$\mathcal{E}$ &	$4.35^{-14} $&	$9.43^{-14} $&	42.77 &	0 &	37.6 \%&	0.0558 &	1 &	24.1 & $G$\\
&$\mathcal{P^+}$ &	$3.18^{-12} $&	$2.86^{-12} $&	32.6 &	0 &	107.5 \%&	0.0428 &	1 &	30.2 & $G$\\
&$\mathcal{A}$ &	$2.34^{-16} $&	$4.52^{-15} $&	25.44 &	34.64 &	4.8 \%&	0.16 &	1 &	11.8 & $G_\mathrm{sym}$\\
\hline\multirow{4}{*}{\rotatebox[origin=c]{90}{Ceres}}
&$\mathcal{M}$ &	$9.87^{-19} $&	$4.42^{-17} $&	22.8 &	37.41 &	2.1 \%&	0.0194 &	0.105 &	1 & $G_\mathrm{sym}$\\
&$\mathcal{E}$ &	$1.19^{-15} $&	$4.36^{-15} $&	35.94 &	1.765 &	25.4 \%&	0.00231 &	0.0415 &	1 & $G$\\
&$\mathcal{P^+}$ &	$8.70^{-14} $&	$9.87^{-14} $&	36.3 &	1.781 &	80.3 \%&	0.00141 &	0.0331 &	1& $G$ \\
&$\mathcal{A}$ &	$3.21^{-17} $&	$3.80^{-16} $&	35.15 &	1.781 &	7.9 \%&	0.0135 &	0.0845 &	1& $G$ \\
\enddata




\end{deluxetable}

\section{Discussion}
\subsection{Meteoroid model}\label{SEC:Discussion_MetModel}
There are several uncertainties that stem from the meteoroid model. Firstly, the meteoroid mass flux at Earth has an intrinsic uncertainty of about 50-60\% based on the latest estimates \citep{CarrilloSanchez_etal_2016, CarrilloSanchez_etal_2020}. This uncertainty linearly scales all quantities in this paper for all three airless bodies discussed here. Secondly, the collisional lifetimes used in our meteoroid model are also subject to uncertainty as discussed for example in \citet{Pokorny_etal_2018} and \citet{Pokorny_etal_2019} for Mercury and Moon, respectively. We \replaced{tested}{analyze} different values of the collisional lifetime multiplier $F_\mathrm{coll} \in [10,50]$ to test the model sensitivity and compare it to the settings used in \citet{Pokorny_etal_2018,Pokorny_etal_2019}. Due to Ceres' proximity to MBA and JFC source populations, the variations of the collisional lifetime has a negligible effect on the model results. The effect on the long-period comet meteoroids (HTCs and OCCs) is similar for Mercury, Moon, and Ceres and does not significantly alter the results.
Thirdly, the size-frequency distributions (SFDs) of our model meteoroid populations are poorly constrained due to the fact that JFC meteoroids dominate the inner solar system budget and the direct measurements of different components of the meteoroid complex are extremely rare \citep[see e.g, Section 2.4 in][]{Pokorny_etal_2019}. 
We ran our meteoroid model calculation using a range of mass indices $[3.4,4.6]$ for each meteoroid population separately, to test their sensitivity to different SFDs. 
Since our model is constrained by the meteoroid mass flux at Earth, the changes in the size-frequency distribution produce $<10\%$ changes in the meteoroid mass flux $\mathcal{M}$, meteoroid energy flux $\mathcal{E}$, and ejecta production rate $\mathcal{P}^+$ at Mercury and the Moon, similar to \citet{Pokorny_etal_2018,Pokorny_etal_2019}. However, the SFD has a higher impact on Ceres, where we record up to $46\%$ higher mass flux for a shallower SFD $\alpha=-3.4$ as compared to our nominal model $\alpha=-4.0$. On the other hand, steeper SFDs result in smaller mass flux by up to $25\%$ for $\alpha=-4.6$. The highest sensitivity to SFD stems from the main-belt meteoroids, due to their extreme proximity to Ceres and their high intrinsic collision probability with Ceres without the need for any dynamical evolution. 

Unlike $\mathcal{M}$, $\mathcal{E}$, and $\mathcal{P}^+$, the area produced by meteoroid bombardment $\mathcal{A}$ and consequently the e-folding lifetime $\mathcal{T}_\mathcal{A}$ do not scale linearly with the meteoroid mass; they scale as $M^{2/3}$. 
\added{The meteoroid model we use here is scaled such that the mass flux at Earth is held constant (see Table \ref{TABLE:Model_Description}). For this reason, $\mathcal{A}$ and $\mathcal{T}_\mathcal{A}$ are more sensitive to the the SFD setting than our other variables}
, which we see for all three airless bodies studied here. The values of $\mathcal{A}$ are approximately two-times larger for $\alpha=-4.6$ compared to our nominal SFD $\alpha=-4.0$, while for the shallow SFD $\alpha=-3.4$ the values are two-times smaller compared to $\alpha=-4.0$. This means, that for steeper SFDs the smaller particles dominate the total impactor cross-section area, while this value is attenuated for shallower SFDs. This impacts also the e-folding lifetime $\mathcal{T}_\mathcal{A}$, which is $\sim 2\times$ shorter for steeper SFD $\alpha=-4.6$ and $\sim 2\times$ longer for shallower SFD $\alpha=-3.4$ as compared to values for $\alpha=-4.0$. The intermediate values of SFD indices fall between the two extremes presented here.

\added{Our meteoroid model represents a range of meteoroids with diameters between $10~\mu$m and $2,000~\mu$m. Particles smaller than $D=10~\mu$m exist in the solar system and add mostly to the number flux experienced by various bodies in the solar system. Their mass flux is small compared to our modeled sample due to their shallower SFD driven by the Poynting-Robertson drag \citep[for SFD at Earth see][] {Love_Brownlee_1993}. Meteoroids with $D\leq1~\mu m$ are effectively blown out of the solar system via radiation pressure \citep{burns_etal_1979} and do not significantly contribute to the quantities analyzed here. By expanding our model to smaller sizes, the results in this article would not change significantly, because our meteoroid model is scaled to provide a certain mass flux at Earth \citep{CarrilloSanchez_etal_2016}. The influence of meteoroids larger than those in our population $D>2,000~\mu$m is more difficult to estimate. We expect meteoroids of $D=2,000~\mu$m to dynamically resemble larger meteoroids because the Poynting-Robertson drag magnitude decreases with increasing particle diameter. This consequently increases the dynamical timescales of larger meteoroids making their dynamical evolution similar to their parent bodies. Asteroidal impacts at Ceres are a focus of \citet{Marchi_etal_2016}. We are not aware of any study that would deal with impacts of comets at Ceres.}

\subsection{Scaling of impact processes}
\label{SEC:Discussion_Impacts}
\added{
In Section \ref{SEC:Definitions}, we establish that the crater-to-projectile cross-section ratio is $F_\mathrm{cr}=63$ using the \citet{Koschny_Grun_2001} experimental results. This experiment showed that the crater diameters were not correlated with impactor velocity, thus the velocity part is missing in Eq. \ref{EQ:Area_Coverage}. On the other hand, a literature overview of impact processes at larger sizes and impact velocities $<8$ \KMS{} shows there is a strong correlation between the impactor velocity and crater size \citep{Holsapple_1993}. 

In order to quantify differences between our estimates for the crater area production rate and scaling laws shown for larger and slower impactors than those we analyze in this article, we use the cratering volume $V_\mathrm{cr}$ estimates from Table 1 in \citet{Holsapple_1993}. We use the strength regime for our impact estimates because our impactors are smaller than 1 mm in radius. Assuming that all meteoroid-induced craters are spherical caps, we get for the volume of the crater
\begin{equation}
    V_\mathrm{cr} = \frac{\pi}{6}h(3a^2+h^2),
\end{equation}
where $a$ is the crater radius, $h$ is the crater depth. Assuming a crater radius-to-depth ratio of $\delta = a/h = 2.55$, $V_\mathrm{cr}$, we get a simple form that allows us to estimate the crater radius $a$:
\begin{equation}
    V_\mathrm{cr}=\frac{\pi a}{6 \delta }(3a^2+\frac{a^2}{\delta^2}) = a^3 \left[\frac{\pi}{6\delta^3}(3\delta^2+1)\right] = a^3 \mathcal{K} \rightarrow a = \left(\frac{V_\mathrm{cr}}{\mathcal{K}}\right)^{1/3}, \mathcal{K} = \left[\frac{\pi}{6\delta^3}(3\delta^2+1)\right].
\end{equation}

\citet{Holsapple_1993} provides a general relation for the crater volume and the impactor characteristics:
\begin{equation}
    V_\mathrm{cr} = C_1 m_\mathrm{met} V_\mathrm{imp}^{\epsilon} = C_1 \frac{\pi}{6}d_\mathrm{met}^3 \rho_\mathrm{met}V_\mathrm{imp}^{\epsilon},
\end{equation}
where $C_1$ is a material constant, $m_\mathrm{met}$ is the impactor/meteoroid mass, $d_\mathrm{met}$  is the impactor/meteoroid diameter, $\rho_\mathrm{met}$ is the impactor/meteoroid bulk density, and $\epsilon$ is the velocity power index. Finally, the crater cross-section $A$ is:
\begin{equation}
    A = \pi a^2 = \pi \left(\frac{V_\mathrm{cr}}{\mathcal{K}}\right)^{2/3} = \frac{\pi d_\mathrm{met}^2}{4} \left[ \frac{4 \pi C_1}{3 \mathcal{K}} \rho_\mathrm{met} \right]^{2/3} v_\mathrm{imp}^{2\epsilon/3} =  \frac{\pi d_\mathrm{met}^2}{4} F_\mathrm{cr}.
    \label{EQ:Holsapple_FINAL}
\end{equation}
For simplicity we assume $\rho_\mathrm{met} = 2000$ kg m$^{-3}$, the value we use in all meteoroid models in this article. From Eq. \ref{EQ:Holsapple_FINAL} we see that $F_\mathrm{cr}$, the crater-to-impactor cross-section ratio, is a function of impact velocity. We recalculate the values of the meteoroid-induced crater cross-section $\mathcal{A}$ for Mercury, the Moon, and Ceres that we show in Section \ref{SEC:Comparison} for two different surface compositions:  dry soil and soft rock\citep[Table 1 in][]{Holsapple_1993}, and summarize our results in Table \ref{TABLE:Holsapple}.

The influence of the impact velocity is the most important for Mercury, where the meteoroids that follow the \citet{Holsapple_1993} formulas are expected to produce on average 23.49 times larger area of craters on soft rock and 10.35 times larger area on dry soil. Then for the soft rock surface, we would expect that the mean surface e-folding time on Mercury is $\overline{\mathcal{T}_\mathrm{A}}=730$ years. Note that formulas from \citet{Holsapple_1993} are not supported by experiments for the size and velocity regimes that we analyze in this article. All three airless bodies that we analyze here are commonly bombarded by meteoroids with much higher impact velocities and smaller sizes than those used in laboratory experiments. Our original decision to use \citet{Koschny_Grun_2001} is based on the fact that this experiment is the closest in impactor size and speed to the meteoroids that we model. 

\citet{Kato_etal_1995} present the results of an experiment with $15 \times 10$ mm cylinder impactors made of water-ice, polycarbonate, aluminium, and basalt. They test two types of targets: ice block and ground snow powder, with a maximum impact velocity of 1 \KMS{}. This experiment uses an order of magnitude larger impactors than the largest meteoroid we model, and velocities much lower than those we record in our model.

The \citet{Shrine_etal_2002} experiments showed that impact cratering of polycrystaline ice by 1-mm aluminium spheres depends heavily on impactor velocity/kinetic energy. Their maximum impact velocity was 7.34 \KMS{}. The entire sample consisted of 16 shots with velocities between 1.07 and 7.34 \KMS{}.

\citet{Sommer_etal_2013} showed a summary of previous laboratory experiments and added impacts of iron meteorite and steel projectiles with velocities between 2.5 and 5.3 \KMS{} onto dry and wet sandstone. None of the lab experiments except for \citet{Koschny_Grun_2001} recodred impact velocities larger than 7 \KMS{} and particle diameters below 800 $\mu$m. These three works show that the impact cratering regime that the meteoroids we model in this article experience, is very poorly constrained by laboratory experiments.

A possible solution to the lack of experimental records is extensive numerical modeling using state-of-the-art Hydrocodes \citep{Elbeshausen_etal_2009, Kraus_etal_2011, Stickle_etal_2020}. However, in this case, we are not aware of any work that analyzes impacts and cratering of micron-sized meteoroids onto regolith or water-ice. }

\begin{deluxetable}{crr|p{0.65in}p{0.65in}p{0.65in}}

\tablecaption{\label{TABLE:Holsapple} Estimates of the average enhancement factor of meteoroid-induced crater cross-sections $\mathcal{A}$ using the \citet{Holsapple_1993} formulas (column 3) with respect to our default value $F_\mathrm{cr} = 63$ for two different surface compositions. Introduction of the impact velocity factor increases the average crater area produced by meteoroids by a factor of 23.49 assuming soft rock surface on Mercury as well as by a factor of $>4.74$ for any combination of surface type and airless body. This would lead to a significant decrease of e-folding lifetimes $\mathcal{T}_A$ for all three airless bodies that we analyze here.}

\tablenum{4}
\tablehead{
\colhead{Surface} & $C_1$	 & \colhead{Volume formula} & \multicolumn{3}{|c}{Average Enhancement Factor}	  \\
type & ~ & ~ & \hfil Mercury & \hfil Moon & \hfil Ceres
} 

\startdata
Dry soil& 0.04 & 	$0.04~m V_\mathrm{imp}^{1.23}$&	\hfil10.35&	\hfil6.33&	\hfil4.74\\
Soft rock& 0.009 &	$0.009~m V_\mathrm{imp}^{1.65}$&	\hfil23.49&	\hfil12.30&	\hfil8.33
\enddata
\end{deluxetable}

\subsection{Axial tilt of Ceres}
The effects of the axial tilt are negligible when assumed over time frames longer than one orbit. This is because the meteoroids impacting the surface of Ceres come from a broad range of ecliptic longitudes and latitudes, as opposed to the Sun, which is represented by a singular point on the celestial sphere. We \replaced{tested}{test} the effects of the axial tilt for the range of $[0^\circ,20^\circ]$, similar to the range of axial tilts shown in \citet{Ermakov_etal_2017}, and \replaced{recorded}{record} $<5\%$ differences between the extreme values of all quantities analyzed in this work. 

\section{Conclusions}
We present the first \deleted{full-fledged} model for micro-meteoroid bombardment effects on the dwarf planet Ceres. Using a detailed shape model, efficient ray-tracing code, and a widely-accepted meteoroid population model for the diameter range of 0.01--2~mm, we estimate the effects of the meteoroid bombardment on the entire Cererean surface analyzed over one precession cycle (Section \ref{Sec:Preccesion_Cycle}), and over one current orbital period (Section \ref{SEC:One_Period}). Finally, the effects meteoroid bombardment experienced by Ceres are compared to those on Mercury and the Moon (Section \ref{SEC:Comparison}).

Here, we summarize the most important findings:
\begin{itemize}
    \item There are no permanently shadowed regions with respect to meteoroid bombardment. The local topography creates up to 80\% difference between occluded and exposed regions (floor vs. rim)
    \item The equatorial regions are producing on average $80\%$ more ejecta than the polar regions, whereas the mass flux is more concentrated at the Cererean poles. However, the mass flux is almost uniform over the entire surface with only $2\%$ variations. The surface e-folding lifetimes are $\sim8\%$ shorter at the Cererean poles as compared to equator.
    \item All areas showing detections of exposed \HHO from \citet{Combe_etal_2019} and all areas of interest from \citet{Ermakov_etal_2017} experience similar rates of meteoroid bombardment (Table \ref{TABLE:Area_Description}). Meteoroids smaller than 2~mm generate orders of magnitude less ejecta from patches of exposed ice than would be required to sustain the water exosphere observed by \citet{Kuppers_etal_2014} and are much lower than estimated surface water-ice sublimation rates. The surface turnover rate is expected to be 1.25~Myr, much longer than the period of obliquity cycles.
    \item Ceres currently experiences a factor of 3--7 seasonal variations in mass flux, energy flux, and ejecta production along its current orbit. This is due to its inclined orbit that takes the dwarf planet far from the invariable plane, i.e., away from the densest parts of the Zodiacal cloud.
    \item Ceres experiences a $\sim10 \times$ and $\sim50\times$ smaller mass flux than the Moon and Mercury, respectively. The meteoroid energy flux and ejecta production rate differences are significantly enhanced by higher impact velocities on the Moon and Mercury resulting in $\sim30 \times$ and $\sim600 \times$ larger effects on the Moon and Mercury, respectively (Table \ref{TABLE:Comparison_Mer_Moon_Ceres}). The same area on Mercury is covered by primarily meteoroid-induced craters 74 times faster than the equivalent area on Ceres, while the lunar surface is covered on timescales 12 times shorter than at Ceres.
\end{itemize}

\acknowledgements
{
P.P. would like to acknowledge the support of the NASA ISFM EIMM,  Planetary Geodesy work packages, NASA awards number 80GSFC17M0002 and 80GSFC21M0002, and Grant Agency of the Czech Republic, grant number: 20-10907S. 
}

\software{
GitHub repository (\url{https://github.com/McFly007/AstroWorks/tree/master/Pokorny_etal_2020_Ceres})
$\bullet$ \texttt{gnuplot} (\url{http://www.gnuplot.info}) $\bullet$
\texttt{tinyobjloader} (\url{https://github.com/tinyobjloader/tinyobjloader}) $\bullet$
\texttt{fastbvh} (\url{https://github.com/brandonpelfrey/Fast-BVH}) $\bullet$
\texttt{swift} \citep{Levison_Duncan_2013} $\bullet$
\texttt{SciPy} (\url{https://www.scipy.org})
}

\bibliography{Papers}{}
\bibliographystyle{aasjournal}

\end{document}